%% file: main.tex
\pdfoutput=1
\documentclass[sigconf,screen]{acmart}

\newif\ifarxiv\arxivfalse

\input{misc/packages}

\input{misc/commands}
\input{misc/config}

\setcopyright{cc}
\setcctype{by}
\acmDOI{10.1145/3832783.3834337}
\acmYear{2026}
\copyrightyear{2026}
\acmISBN{979-8-4007-2882-2/2026/10}
\acmConference[ASE '26]{Proceedings of the 41st IEEE/ACM International Conference on Automated Software Engineering}{October 12--16, 2026}{Munich, Germany}
\acmBooktitle{Proceedings of the 41st IEEE/ACM International Conference on Automated Software Engineering (ASE '26), October 12--16, 2026, Munich, Germany}
\acmSubmissionID{ase26main-p90-p}
\received{2026-03-26}
\received[accepted]{2026-06-18}

\begin{document}

\title{LLM-Based Agents for Identifying Bug-Introducing Commits}

\author{Niklas Risse}
\correspondingauthor
\orcid{0000-0002-0666-5025}
\affiliation{%
  \institution{MPI-SP and CISPA}
  \country{Germany}
}
\email{niklasrisse@gmail.com}

\author{Marcel Böhme}
\orcid{0000-0002-4470-1824}
\affiliation{%
  \institution{MPI-SP and CISPA}
  \country{Germany}
}
\email{marcel.boehme@acm.org}

\begin{abstract}
  \input{sections/abstract}
\end{abstract}



\maketitle

\input{sections/introduction}

\input{sections/background}

\input{sections/part_1}

\input{sections/part_2}

\input{sections/discussion}

\input{sections/threats_to_validity}

\input{tables/table_effectiveness_file_combined}
\input{tables/table_effectiveness_function_combined}
\input{tables/table_cheaper_models_split}

\input{sections/conclusion}

\appendix

\input{sections/appendix}

\ifarxiv\else
\input{sections/data_availability}
\fi

\input{sections/acknowledgements}

\balance
\bibliographystyle{ACM-Reference-Format}
\bibliography{references}

\end{document}

%% file: misc/packages.tex
\usepackage[table]{xcolor}
\usepackage{colortbl}
\usepackage{booktabs}
\usepackage{lipsum}
\usepackage[most]{tcolorbox}
\usepackage{enumitem}
\usepackage{tikz}
\usepackage{pgfplots}
\pgfplotsset{compat=1.18}
\usepackage{makecell}
\usepackage{subcaption}
\usepackage{pifont}
\usepackage{xparse}
\usepackage{xspace}
\usepackage{balance}
\usetikzlibrary{shapes.geometric, arrows.meta, positioning, fit, backgrounds, calc}

\definecolor{inputgreen}{HTML}{2E7D32}
\definecolor{processblue}{HTML}{1565C0}
\definecolor{agentpurple}{HTML}{7B1FA2}
\definecolor{decisionorange}{HTML}{EF6C00}
\definecolor{outputred}{HTML}{C62828}
\definecolor{stage1bg}{HTML}{F5F5F5}
\definecolor{stage2bg}{HTML}{F5F5F5}
\definecolor{arrowgray}{HTML}{546E7A}
\definecolor{textgray}{HTML}{37474F}

%% file: misc/commands.tex
\newtcolorbox{rqbox}{
  colframe=black,colback=white,boxrule=0.5pt,arc=4pt,
  left=6pt,right=6pt,top=6pt,bottom=6pt,boxsep=0pt,width=\columnwidth
}

\newcommand{\boldpar}[1]{\smallskip\noindent\textbf{#1.}\xspace}

\newcommand{\result}[1]{%
\begin{tcolorbox}[colframe=black,boxrule=0.5pt,arc=4pt,
      left=6pt,right=6pt,top=6pt,bottom=6pt,boxsep=0pt,width=\columnwidth]%
      {\emph{#1}}
\end{tcolorbox}%
}

%% file: misc/config.tex
\newcommand{\artifacturl}{https://github.com/niklasrisse/agents-for-szz}

%% file: sections/abstract.tex
Śliwerski, Zimmermann, and Zeller (SZZ) just won the 2026 ACM SIGSOFT Impact Award for asking:

\begin{center}
    When do changes induce fixes?
\end{center}

Their paper from 2005 served as the foundation for a wide array of approaches aimed at identifying bug-introducing changes (or commits) from fix commits in software repositories. But even after two decades of progress, the best-performing approach from 2025 yields a modest increase of 10 percentage points in F1-score on the most popular Linux kernel dataset.

In this paper, we uncover how and why LLM-based agents can substantially advance the state-of-the-art in identifying bug-introducing commits from fix commits. We propose a simple agentic workflow based on searching a set of candidate commits and find that it raises the F1-score from 0.64 to 0.81 on the most popular Linux kernel dataset, a bigger jump than between the original 2005 method (0.54) and the previous SOTA (0.64). We also uncover \emph{why} agents are so successful: They derive short \emph{greppable} patterns from the fix commit diff and message and use them to effectively search and find bug-introducing commits in large candidate sets. Finally, we also discuss how these insights might enable further progress in root cause understanding and repair.

%% file: sections/introduction.tex
\section{Introduction}
The problem of identifying bug-introducing commits from fix commits has seen remarkable interest from the software engineering research community, demonstrated by the more than 1300 citations of the award-winning paper by Śliwerski, Zimmermann, and Zeller (SZZ)~\cite{sliwerski_2005_original_szz}. The problem is also highly relevant for practitioners, since identifying bug-introducing changes is essential to determine which software versions are affected by security-critical bugs~\cite{chen_2025_how_far_are_we}.

Even though many new approaches have been released since 2005~\cite{sunghun_2006_agszz, davies_2014_lszz_rszz, dacosta_2017_maszz, neto_2018_raszz, bao_2022_vszz, tang_2023_neuralszz, tang_2025_llm4szz}, the performance gains remain fairly limited~\cite{chen_2025_how_far_are_we}. The foundational 2005 SZZ paper~\cite{sliwerski_2005_original_szz} already achieves 0.54 F1-score on a high-quality developer-annotated dataset derived from the Linux kernel~\cite{Lyu_2024_Linux_Kernel_Dataset}. On the same dataset, the state-of-the-art method from 2025~\cite{tang_2025_llm4szz} scores just 10 percentage points higher, with an F1-score of 0.64. To break the trend of small incremental improvements, entirely new approaches are needed. 

In this paper, we study LLM-based agents for the task of identifying bug-introducing commits (BICs) from fix commits.

\boldpar{Part 1: SZZ-Agent} First, we introduce \emph{SZZ-Agent}, which targets fix commits where the original SZZ algorithm struggles: cases where SZZ finds no candidate BICs (e.g., because the fix adds code rather than removing it), or where its prediction is likely wrong. SZZ-Agent collects candidate BICs from the file history and binary searches over them, examining the source code at each version to determine when the bug first appeared.

In our evaluation, SZZ-Agent convincingly beats all previous approaches, raising the state-of-the-art score by 9 to 13 percentage points on four popular C/C++ and Java datasets. Beyond effectiveness, we dive into \emph{why} SZZ-Agent works using five ablation studies to investigate whether (a) the performance comes from data leakage, (b) what context SZZ-Agent relies on, (c) whether the performance gain can be explained by a better LLM-backbone, (d) what components of SZZ-Agent contribute the most to its success, and (e) how a specific hyperparameter, the commit selection threshold, controls its performance.

\boldpar{Part 2: Simple-SZZ-Agent} During our ablation studies, we discover that SZZ-Agent can be drastically simplified: skipping binary search entirely and letting the agent directly select from all candidates yields the best F1-score at 43\% lower LLM token cost (in USD) per fix commit. Based on this finding, we propose Simple-SZZ-Agent, which collects candidates from the file history and directly asks the agent to select the bug-introducing commit---yet it outperforms SZZ-Agent. Surprisingly, it is also highly scalable: its resource consumption (e.g., tokens) is almost independent of the number of candidate commits. To understand why, we investigated what the agent actually does.

This led us to uncover \emph{why} Simple-SZZ-Agent works so well: The agent distills the fix commit message and diff into short string patterns, greps the candidate set for matches, and then reads and evaluates the results, repeating this process until it identifies the bug-introducing commit.

To finalize our study, we propose directions for future work based on our findings. The agent's ability to compress knowledge about a bug into short greppable patterns opens avenues beyond identifying bug-introducing commits: these patterns could help practitioners understand how a bug was introduced, detect similar bugs in other code, or provide context for generating fixes.

\begin{figure*}[t]
    \centering
    \resizebox{\textwidth}{!}{\input{figures/approach_overview}}
    \caption{SZZ-Agent: Stage 1 applies standard SZZ to the deleted lines and uses an agent to select BICs from the candidates. Stage 2 activates when Stage 1 fails: it collects file histories, binary searches over commits by having an agent analyze the source code at each midpoint, and narrows the search space until the remaining candidates are few enough for direct selection.}
    \label{fig:szz-agent-overview}
\end{figure*}
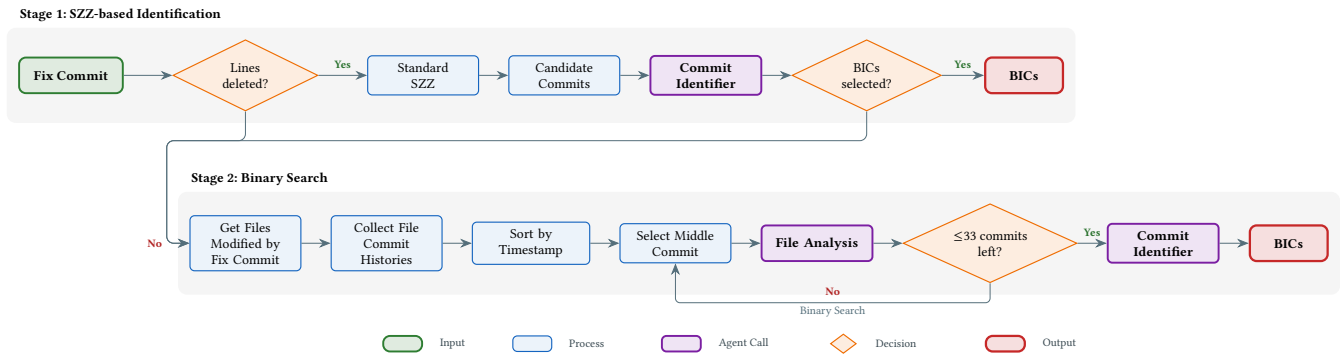

\boldpar{Contributions} We make the following key contributions:
  \begin{itemize}
      \item \textbf{SZZ-Agent:} An agentic workflow that uses binary search over source code versions to identify bug-introducing commits, outperforming all previous approaches by at least 9 percentage points in F1-score on four popular datasets. We conduct five ablation studies to understand why it works.
      \item \textbf{Simple-SZZ-Agent:} A simpler agent discovered during our ablations that outperforms even SZZ-Agent by directly selecting the bug-introducing commit from a large candidate set. We uncover its strategy: it compresses the fix commit into short greppable patterns to efficiently search the candidates.
  \end{itemize}

%% file: figures/approach_overview.tex
\begin{tikzpicture}[
    node distance=0.4cm and 0.5cm,
    >={Stealth[length=1.8mm, width=1.4mm]},
    inputnode/.style={
        rectangle, rounded corners=3pt,
        minimum width=1.6cm, minimum height=0.55cm,
        text centered, draw=inputgreen, line width=0.85pt,
        fill=inputgreen!15, font=\scriptsize\bfseries
    },
    procnode/.style={
        rectangle, rounded corners=2pt,
        minimum width=1.6cm, minimum height=0.55cm,
        text centered, draw=processblue, line width=0.5pt,
        fill=processblue!8, font=\scriptsize,
        text width=1.5cm, align=center
    },
    agentnode/.style={
        rectangle, rounded corners=2pt,
        minimum width=1.6cm, minimum height=0.55cm,
        text centered, draw=agentpurple, line width=0.85pt,
        fill=agentpurple!12, font=\scriptsize\bfseries,
        text width=1.5cm, align=center
    },
    decnode/.style={
        diamond, aspect=2.2,
        minimum width=2.2cm,
        minimum height=1.1cm,
        text centered, draw=decisionorange, line width=0.5pt,
        fill=decisionorange!12, font=\scriptsize,
        inner sep=3pt,
        align=center
    },
    outnode/.style={
        rectangle, rounded corners=3pt,
        minimum width=1.2cm, minimum height=0.55cm,
        text centered, draw=outputred, line width=1pt,
        fill=outputred!15, font=\scriptsize\bfseries
    },
    myarrow/.style={->, line width=0.42pt, draw=arrowgray},
    yeslbl/.style={font=\tiny\bfseries, text=inputgreen!90!black},
    nolbl/.style={font=\tiny\bfseries, text=outputred!90!black}
]


\node[inputnode] (start) {Fix Commit};

\node[decnode, right=0.75cm of start] (dec1) {Lines\\deleted?};

\node[procnode, right=0.75cm of dec1] (szz) {Standard\\SZZ};

\node[procnode, right=0.45cm of szz] (cand1) {Candidate\\Commits};

\node[agentnode, right=0.45cm of cand1] (cia1) {Commit\\Identifier};

\node[decnode, right=0.45cm of cia1] (dec2) {BICs\\selected?};

\node[outnode, right=0.65cm of dec2] (out1) {BICs};


\node[procnode, below=1.6cm of dec1] (getfiles) {Get Files \\ Modified by \\ Fix Commit};

\node[procnode, right=0.45cm of getfiles] (gethist) {Collect File\\Commit Histories};

\node[procnode, right=0.45cm of gethist, text width=1.6cm, minimum width=1.7cm] (sortc) {Sort by\\Timestamp};

\node[procnode, right=0.45cm of sortc] (selectmid) {Select Middle\\Commit};

\node[agentnode, right=0.45cm of selectmid] (faa) {File Analysis};

\node[decnode, right=0.45cm of faa] (dec3) {$\leq$33 commits\\left?};

\node[agentnode, right=0.45cm of dec3] (cia2) {Commit\\Identifier};

\node[outnode, right=0.45cm of cia2] (out2) {BICs};


\begin{scope}[on background layer]
    \node[fit=(start)(dec1)(szz)(cand1)(cia1)(dec2)(out1),
          fill=stage1bg, rounded corners=4pt, inner sep=5pt] (bg1) {};
    \node[fit=(getfiles)(gethist)(sortc)(selectmid)(faa)(dec3)(cia2)(out2),
          fill=stage2bg, rounded corners=4pt, inner sep=5pt] (bg2) {};
\end{scope}


\node[font=\scriptsize\bfseries, text=black, anchor=north west]
    at ($(bg1.north west) + (0.1, +0.4)$) {Stage 1: SZZ-based Identification};
\node[font=\scriptsize\bfseries, text=black, anchor=north west]
    at ($(bg2.north west) + (0.1, +0.4)$) {Stage 2: Binary Search};


\draw[myarrow] (start) -- (dec1);
\draw[myarrow] (dec1) -- node[yeslbl, above] {Yes} (szz);
\draw[myarrow] (szz) -- (cand1);
\draw[myarrow] (cand1) -- (cia1);
\draw[myarrow] (cia1) -- (dec2);
\draw[myarrow] (dec2) -- node[yeslbl, above] {Yes} (out1);


\draw[myarrow, rounded corners=3pt]
    (dec1.south) -- ++(0,-0.45) -| ($(getfiles.west) + (-0.35, 0)$) -- (getfiles.west)
    node[nolbl, left, pos=0.12] {No};

\draw[myarrow, rounded corners=3pt]
    (dec2.south) -- ++(0,-0.45) -| ($(getfiles.west) + (-0.35, 0)$) -- (getfiles.west);


\draw[myarrow] (getfiles) -- (gethist);
\draw[myarrow] (gethist) -- (sortc);
\draw[myarrow] (sortc) -- (selectmid);
\draw[myarrow] (selectmid) -- (faa);
\draw[myarrow] (faa) -- (dec3);
\draw[myarrow] (dec3) -- node[yeslbl, above] {Yes} (cia2);
\draw[myarrow] (cia2) -- (out2);

\draw[myarrow, rounded corners=2pt]
    (dec3.south) -- ++(0,-0.3) -| node[nolbl, above, pos=0.25] {No} (selectmid.south);

\node[font=\tiny, text=arrowgray, anchor=north] at ($(selectmid.south)!0.5!(dec3.south) + (0, -0.4)$) {Binary Search};


\begin{scope}[shift={($(bg2.south) + (0, -0.75)$)}]
    \node[inputnode, minimum width=0.6cm, minimum height=0.24cm, font=\tiny\bfseries, inner sep=0.3pt] at (-5.5,0) (L1) {};
    \node[font=\tiny, text=textgray, anchor=west] at ($(L1.east)+(0.15,0)$) {Input};

    \node[procnode, minimum width=0.6cm, minimum height=0.24cm, font=\tiny, inner sep=0.3pt, text width=0.4cm] at (-3.5,0) (L2) {};
    \node[font=\tiny, text=textgray, anchor=west] at ($(L2.east)+(0.15,0)$) {Process};

    \node[agentnode, minimum width=0.6cm, minimum height=0.24cm, font=\tiny\bfseries, inner sep=0.3pt, text width=0.4cm] at (-1.2,0) (L3) {};
    \node[font=\tiny, text=textgray, anchor=west] at ($(L3.east)+(0.15,0)$) {Agent Call};

    \node[decnode, minimum width=0.4cm, minimum height=0.3cm, aspect=1.6, font=\tiny, inner sep=0pt] at (1.3,0) (L4) {};
    \node[font=\tiny, text=textgray, anchor=west] at ($(L4.east)+(0.18,0)$) {Decision};

    \node[outnode, minimum width=0.6cm, minimum height=0.24cm, font=\tiny\bfseries, inner sep=0.3pt] at (3.8,0) (L5) {};
    \node[font=\tiny, text=textgray, anchor=west] at ($(L5.east)+(0.15,0)$) {Output};
\end{scope}

\end{tikzpicture}

%% file: sections/background.tex
\section{Background}
\label{sec:background}

\boldpar{Problem Definition} Given a fix commit $c_{\text{fix}}$ that repairs a bug in a software repository, the goal is to identify the set of bug-introducing commits $B \subseteq H$, where $H$ is the repository history prior to $c_{\text{fix}}$. Each commit $b \in B$ introduced a change that caused or contributed to the bug that $c_{\text{fix}}$ resolves. In practice, $|B|$ is often one but can be larger when a bug arises from changes spread across multiple commits.

\subsection{Related Work}

\boldpar{The SZZ Algorithm} The original SZZ algorithm~\citep{sliwerski_2005_original_szz} identifies bug-introducing commits by tracing the lines deleted in a fix commit back to the commits that last modified them using version control annotations. These earlier commits form the set of bug-introducing candidates. While simple, this approach rests on the assumption that the lines removed by the fix are the same lines that introduced the bug. This assumption does not always hold: for instance, a commit may introduce a function that fails to validate its inputs, yet the fix adds a validation check rather than modifying the original function code. In such cases, the deleted lines in the fix are unrelated to the actual bug-introducing change. More generally, the approach suffers from false positives because not every deleted line is related to the bug fix, and it cannot identify bug-introducing commits for fixes that only add lines without deleting any, since there are no lines to trace back.

\boldpar{Filtering-based Refinements} Subsequent work tries to overcome the limitations of SZZ by refining how candidates are generated, filtered, or selected. AG-SZZ~\citep{sunghun_2006_agszz} uses annotation graphs and excludes non-semantic changes such as comments and formatting. MA-SZZ~\citep{dacosta_2017_maszz} further excludes meta-changes such as modifications to module declarations and import statements. RA-SZZ~\citep{neto_2018_raszz} integrates refactoring detection tools to remove behavior-preserving changes, which account for 6.5\% of changes incorrectly flagged as bug-introducing. L-SZZ and R-SZZ~\citep{davies_2014_lszz_rszz} reduce the candidate set to a single commit: L-SZZ selects the candidate that changes the largest number of lines, while R-SZZ selects the most recent candidate. V-SZZ~\citep{bao_2022_vszz} targets security vulnerabilities specifically by recursively tracing lines back to the earliest commit that modified them, based on the observation that vulnerabilities are often introduced in early versions of the code.

\boldpar{Learning-based Approaches} More recent methods apply machine learning to refine which lines are traced or to assess candidates directly. Neural-SZZ~\citep{tang_2023_neuralszz} uses a graph neural network to rank the deleted lines in the fix commit by bug-relevance, then applies SZZ only to the top-ranked lines. LLM4SZZ~\citep{tang_2025_llm4szz}, published at ACM ISSTA 2025 and to our knowledge the current state-of-the-art technique, uses two LLM-based strategies: rank-based identification, which selects and ranks buggy statements in the fix commit, and context-enhanced identification, which prompts the LLM with contextual information about the fix commit and each SZZ candidate to assess whether a candidate truly introduced the bug.

\result{All of these previous approaches remain critically limited by their performance. Despite two decades of refinements, the cumulative improvement over the original SZZ remains limited to 5--10 percentage points in F1-score using the five datasets in our evaluation setup (see \S\ref{sec:szz-agent} and \S\ref{sec:simple-szz-agent}).}

\boldpar{Software Engineering Agents} Agents are software systems that use large language models (LLMs) trained to invoke external tools such as command-line interfaces, file readers, and web search~\citep{schick_2023_toolformer, yao_2023_react} to achieve goals specified by a user or other agents. Agents have proven effective across a range of software engineering tasks, including automated program repair and autonomous issue resolution~\citep{jin_2025_agents_in_se_survey, yang_2024_sweagent, zhang_2024_autocoderover}. Several mature agent platforms exist today: Claude Code~\citep{anthropic_claude_code} is a widely adopted proprietary coding agent, while OpenHands~\citep{wang_2024_openhands} is a popular open-source platform that supports multiple LLM backends. LLM-based agents promise to be more effective than previous approaches at determining when a bug was first introduced, as they can reason about the semantics of code changes~\citep{gu_2024_cruxeval}, navigate repository histories with tools, and examine source code across versions. Whether this promise holds is the central question of this paper.

%% file: sections/part_1.tex
\section{Part 1: SZZ-Agent}
\label{sec:szz-agent}
To test whether the capabilities of LLM-based agents translate into effective bug-introducing commit identification, we design SZZ-Agent, an agentic approach to identify bug-introducing commits. In this section we explain how SZZ-Agent works (\S\ref{sec:szz-agent:approach}) and present our evaluation (\S\ref{sec:szz-agent:experimental_setup}--\S\ref{sec:szz-agent:experiments}).

\subsection{Approach}
\label{sec:szz-agent:approach}
Figure~\ref{fig:szz-agent-overview} shows an overview of SZZ-Agent. It consists of two stages.

\boldpar{Stage 1: SZZ-based Identification} The approach starts from a fix commit, defined by its diff, showing which lines were added and/or deleted, and its commit message. Via standard SZZ, all deleted lines are traced back to the commits that last introduced them, forming the SZZ candidate set. We then provide a LLM-based agent with the fix commit message, the fix commit diff, and for each candidate commit its message and diff (all with five lines of context). The agent is instructed to analyze the fix to understand the bug, examine each candidate for whether it introduced the bug, and select the earliest bug-introducing commit, or \texttt{NONE} if no candidate is responsible.

\boldpar{Stage 2: Binary Search} For entries where Stage 1 abstains or where no SZZ candidates exist, we apply a second stage that operates over the full commit history of every file touched by the fix. We then perform a binary search over this history: at each step, the agent receives the full file contents at the candidate commit, the buggy version immediately before the fix, the fixed version, the fix commit message, and the fix commit diff. Based on this, the agent determines whether the bug is already present at the candidate commit. Once the search window narrows to a tunable threshold of candidates (default: 33; see \S\ref{sec:szz-agent:experiments:rq2e} for an ablation), the approach switches to direct candidate selection, presenting the remaining candidates' diffs and messages for the agent to choose the most likely bug-introducing commit.

\subsection{Experimental Setup}
\label{sec:szz-agent:experimental_setup}
The goal of the evaluation is to investigate how effective SZZ-Agent is and why it works. To achieve this goal, we ask the following research questions:

\begin{rqbox}
\textbf{RQ1} \textbf{(Effectiveness):} How effective is SZZ-Agent at identifying bug-introducing commits?
\end{rqbox}

\begin{rqbox}
\textbf{RQ2} \textbf{(Ablations):} Why is SZZ-Agent effective?\\[2pt]
\hspace{1em}\textbf{(a) Data Leakage:} Can the effectiveness of SZZ-Agent be explained by data leakage?\\
\hspace{1em}\textbf{(b) LLM Backbone:} Is SZZ-Agent more effective than previous state-of-the-art because it uses better LLMs?\\
\hspace{1em}\textbf{(c) Context:} What contextual information enables SZZ-Agent to determine the BICs?\\
\hspace{1em}\textbf{(d) Stage Contribution:} What is the contribution of Stage 1 vs. Stage 2?\\
\hspace{1em}\textbf{(e) Hyperparameter:} How does the performance vary with the threshold for direct commit selection in Stage 2?
\end{rqbox}

The following paragraphs describe the experimental setup shared across all experiments. Experiment-specific details are provided in the individual setup paragraphs in \S\ref{sec:szz-agent:experiments}.

\subsubsection{Datasets and Pre-processing}

\input{tables/table_effectiveness}

We use the most popular datasets across two programming language families, which are also the datasets used in the LLM4SZZ evaluation~\citep{tang_2025_llm4szz}.

\boldpar{DS\_LINUX} Since October 2013, Linux kernel developers label bug-fixing patches with the commit identifiers of the corresponding bug-introducing commits via \texttt{Fixes:} tags. \citet{Lyu_2024_Linux_Kernel_Dataset} leverage this practice to construct a developer-annotated dataset of 76,046 fix commit and bug-introducing commit pairs.

\boldpar{DS\_LINUX-26} To evaluate on fix commits published after the training data cutoff of the LLMs we use, we apply the methodology of \citet{Lyu_2024_Linux_Kernel_Dataset} to collect 5275 fix commit and bug-introducing commit pairs from the Linux kernel for fix commits published between September 1, 2025 and January 31, 2026.

\boldpar{DS\_GITHUB-c and DS\_GITHUB-j} \citet{rosa_2021_ds_github} construct a developer-annotated dataset from GitHub projects by using natural language processing to identify fix commits in which developers explicitly reference the bug-introducing commits, followed by manual filtering. We split this dataset by language into DS\_GITHUB-c (C/C++ projects) and DS\_GITHUB-j (Java projects).

\boldpar{DS\_APACHE} \citet{wen_2019_ds_apache} manually collect and verify the bug-introducing commits of 333 bugs from seven projects of the Apache ecosystem. We use the version of this dataset from the LLM4SZZ evaluation~\citep{tang_2025_llm4szz}, which retains 243 of these bugs.

\boldpar{Pre-processing} To prevent the agents from identifying the bug-introducing commit by matching commit hashes referenced in the fix commit message (e.g., in \texttt{Fixes:} tags), we redact commit hash references from all provided diffs and messages. Specifically, all occurrences of ground-truth commit hashes are replaced with \texttt{COMMIT\_HASH}, and lines containing \texttt{Fixes:} tags that reference these commits are removed entirely. The agents cannot recover this redacted information: they operate on a working directory that contains the relevant files and diffs as separate files rather than the actual git repository, and internet access is blocked.

\subsubsection{Baselines}
We compare against eight SZZ variants described in \S\ref{sec:background}: SZZ, AG-SZZ, L-SZZ, R-SZZ, MA-SZZ, RA-SZZ, V-SZZ, and LLM4SZZ. For all baselines except V-SZZ and LLM4SZZ, we use the implementations provided by \citet{rosa_2023_szzvariants}. For V-SZZ and LLM4SZZ, we use the implementations provided by the respective authors~\citep{bao_2022_vszz, tang_2025_llm4szz}. To the best of our knowledge, LLM4SZZ~\citep{tang_2025_llm4szz}, published at ACM ISSTA 2025, represents the state of the art.

\subsubsection{Agents}
We use Claude Code~\citep{anthropic_claude_code} as our agent framework. Claude Code provides 30 built-in tools~\citep{anthropic_claude_code_tools}, including bash command execution, file reading and editing, glob and grep search, web search, and sub-agent spawning---capabilities that are essential for inspecting source code across repository versions. We block all agents from using web search tools (e.g., WebSearch and WebFetch in Claude Code) to prevent them from searching the internet for solutions. As LLM backbones, we use models from Anthropic and OpenAI; see the individual experiment setups for details on which models we used. We explore open-source agents and open-source LLM backbones in Part~2 (\S\ref{sec:simple-szz-agent:experiments:rq4c}).

\subsubsection{Metrics}

Following \citet{Lyu_2024_Linux_Kernel_Dataset}, we evaluate using precision, recall, and F1-score. Since each fix commit $c_{\text{fix}}$ can have multiple ground-truth bug-introducing commits $B$ and multiple predicted commits $\hat{B}$, we compute per-fix-commit precision and recall as:
\begin{equation}
\text{Precision}(c_{\text{fix}}) = \frac{|B \cap \hat{B}|}{|\hat{B}|}, \quad \text{Recall}(c_{\text{fix}}) = \frac{|B \cap \hat{B}|}{|B|}.
\end{equation}
We then macro-average across all fix commits and compute F1-score as the harmonic mean of the averaged precision and recall.

\subsection{Experiments}
\label{sec:szz-agent:experiments}

\subsubsection{RQ1: Effectiveness}
We evaluate whether SZZ-Agent outperforms existing SZZ variants.

\boldpar{Setup} We run SZZ-Agent with Claude Opus 4.5 as the LLM backbone on all four established datasets: DS\_LINUX (200 samples), DS\_GITHUB-c (100 samples), DS\_GITHUB-j (75 samples, limited by the number of Java entries in the dataset), and DS\_APACHE (all 243 samples). Sample sizes were chosen based on budget considerations; we use the Wilcoxon signed-rank test~\citep{wilcoxon_1945} at $p < 0.01$ to assess statistical significance and report the rank-biserial correlation~\citep{kerby_2014} as effect size. We include the sampling scripts with fixed random seeds for exact reproducibility in our artifact. We compare against all eight baselines described in \S\ref{sec:szz-agent:experimental_setup}.

\boldpar{Results} Table~\ref{tab:results} shows the results. SZZ-Agent outperforms all baselines on every dataset. Compared to LLM4SZZ, the previous state of the art, SZZ-Agent improves F1-score by 13 percentage points on DS\_LINUX (0.77 vs.\ 0.64), 12 on DS\_GITHUB-c (0.75 vs.\ 0.63), 12 on DS\_GITHUB-j (0.67 vs.\ 0.55), and 9 on DS\_APACHE (0.60 vs.\ 0.51). SZZ-Agent achieves the highest precision and recall on all datasets, with the exception of recall on DS\_GITHUB-j and DS\_APACHE, where standard SZZ achieves 0.68 (vs.\ 0.67) and 0.58 (vs.\ 0.56), respectively. All improvements are statistically significant ($p < 0.01$) with large effect sizes, except for DS\_GITHUB-j vs.\ LLM4SZZ where the effect size is large but $p = 0.013$.

\subsubsection{RQ2a: Data Leakage}
\label{sec:szz-agent:experiments:rq2a}
Since the fix commits in DS\_LINUX, DS\_GITHUB-c, DS\_GITHUB-j, and DS\_APACHE were published before the training data cutoff of Claude Opus 4.5 (August 2025~\citep{anthropic_models_overview}), the model may have encountered them during training. We investigate whether SZZ-Agent's effectiveness can be explained by such data leakage.

\boldpar{Setup} We evaluate SZZ-Agent on DS\_LINUX-26, which we collected by applying the same methodology as \citet{Lyu_2024_Linux_Kernel_Dataset} used for DS\_LINUX but on more recent kernel data. It contains 100 randomly sampled fix commits published between September 2025 and January 2026---after the training data cutoff of Claude Opus 4.5.

\boldpar{Results} Table~\ref{tab:results_leakage} shows the results. SZZ-Agent achieves 0.81 F1-score on DS\_LINUX-26, which is even higher than its 0.77 on DS\_LINUX. All improvements over baselines are statistically significant ($p < 0.01$) with large effect sizes. This rules out data leakage as an explanation for SZZ-Agent's effectiveness.

\subsubsection{RQ2b: LLM Backbone}
We investigate whether SZZ-Agent's improvement over LLM4SZZ is simply due to using a more capable LLM.

\boldpar{Setup} We run LLM4SZZ with three different LLM backbones---gpt-4o-mini (its default), gpt-5.2, and Claude Opus 4.5---and compare against SZZ-Agent with Claude Opus 4.5. All experiments use DS\_LINUX (200 samples).

\boldpar{Results} Table~\ref{tab:results_models} shows the results. Upgrading the LLM backbone of LLM4SZZ from gpt-4o-mini to gpt-5.2 or Claude Opus 4.5 does not meaningfully improve its F1-score (0.64 $\rightarrow$ 0.61 $\rightarrow$ 0.65). In contrast, SZZ-Agent with the same Claude Opus 4.5 backbone achieves 0.77. This demonstrates that SZZ-Agent's advantage stems from its agentic approach rather than from using a better LLM.

\result{\textbf{Finding 1:} SZZ-Agent outperforms all previous approaches by at least 9 percentage points in F1-score on four established datasets (RQ1). This improvement cannot be explained by data leakage (RQ2a) or by the use of a more capable LLM backbone (RQ2b), but stems from the agentic approach itself.}

\subsubsection{RQ2c: Context Ablation}
\label{sec:szz-agent:experiments:rq2c}
We investigate which contextual information enables SZZ-Agent to identify bug-introducing commits.

\boldpar{Setup} We ablate the fix commit message and the fix commit diff by selectively withholding them from the agent. We run all variants on DS\_LINUX-26 with Claude Opus 4.5.

\boldpar{Results} Table~\ref{tab:results_without_message_or_diff} shows the results. Without both the commit message and the diff, SZZ-Agent drops to 0.48 F1-score, below standard SZZ (0.56). Providing only the diff recovers performance to 0.73, and providing only the message reaches 0.77---each individually already exceeding the previous state-of-the-art (LLM4SZZ, 0.62 on DS\_LINUX-26). This suggests that the two sources carry partially redundant information. However, combining both achieves 0.81, indicating that each source also provides complementary information not captured by the other.

\result{\textbf{Finding 2:} The fix commit message and the fix commit diff carry partially redundant information: each alone already surpasses the previous state-of-the-art. However, combining both yields the highest performance (0.81 F1), showing that neither source fully subsumes the other.}

\input{tables/table_data_leakage}

\input{tables/table_llm4szz_different_models}

\input{tables/table_without_message_or_diff}

\subsubsection{RQ2d: Stage 1 vs. Stage 2}
We investigate the individual contribution of each stage to the overall pipeline.

\boldpar{Setup} We run Stage~1 and Stage~2 in isolation and compare against the combined pipeline. All experiments use DS\_LINUX-26 with Claude Opus 4.5. We also report the average cost per fix commit in USD.

\boldpar{Results} Table~\ref{tab:stage_01_vs_stage_02} shows the results. Stage~1 alone achieves only 0.48 F1-score at \$0.18 per fix commit, while Stage~2 alone reaches 0.76 at \$1.20. The combined pipeline achieves the best F1-score of 0.81 at \$0.67, demonstrating that both stages are complementary: Stage~1 cheaply resolves straightforward cases, reducing the number of entries that require the more expensive Stage~2.

\result{\textbf{Finding 3:} Stage~1 and Stage~2 are complementary. Stage~1 cheaply resolves straightforward cases, while Stage~2 handles the remaining difficult cases. Their combination achieves the best F1-score at moderate cost.}

\subsubsection{RQ2e: Commit Selection Threshold}
\label{sec:szz-agent:experiments:rq2e}
We investigate how the commit selection threshold---the number of remaining candidates at which Stage~2 switches from binary search to direct selection---affects performance and cost.

\boldpar{Setup} We vary the threshold across values of 8, 32, 128, 512, and $\infty$ (no binary search, always direct selection). All experiments use DS\_LINUX-26 with Claude Opus 4.5.

\boldpar{Results} Table~\ref{tab:commit_identifier_threshold} shows the results. Performance is remarkably stable across thresholds, ranging from 0.76 to 0.82 F1-score. Higher thresholds reduce cost (\$0.68 at threshold~8 vs.\ \$0.33 at $\infty$) because fewer binary search steps are needed. The $\infty$ threshold---which skips binary search entirely and always uses direct selection---achieves the best F1-score (0.82) at the lowest cost (\$0.33).

\result{\textbf{Finding 4 (Most Surprising):} Skipping binary search entirely and always using direct candidate selection yields the best F1-score (0.82) at the lowest cost. The agent can identify the bug-introducing commit from a large candidate set in a single step, without iterative narrowing. This observation led us to question whether binary search is necessary at all, and motivated the development of a drastically simplified approach: Simple-SZZ-Agent (\S\ref{sec:simple-szz-agent}).}

\input{tables/table_stage_01_vs_stage_02}
\input{tables/table_commit_identifier_threshold}

%% file: tables/table_effectiveness.tex
\begin{table*}[t]
\centering
\footnotesize
\caption{RQ1: SZZ-Agent vs.\ baselines across four datasets. Best results per dataset in \textbf{bold}.}
\label{tab:results}
\begin{tabular}{lcccccccccccc}
\toprule
 & \multicolumn{3}{c}{\textbf{DS\_LINUX}} & \multicolumn{3}{c}{\textbf{DS\_GITHUB-c}} & \multicolumn{3}{c}{\textbf{DS\_GITHUB-j}} & \multicolumn{3}{c}{\textbf{DS\_APACHE}} \\
\cmidrule(lr){2-4} \cmidrule(lr){5-7} \cmidrule(lr){8-10} \cmidrule(lr){11-13}
\textbf{Approach} & \textbf{Precision} & \textbf{Recall} & \textbf{F1-Score} & \textbf{Precision} & \textbf{Recall} & \textbf{F1-Score} & \textbf{Precision} & \textbf{Recall} & \textbf{F1-Score} & \textbf{Precision} & \textbf{Recall} & \textbf{F1-Score} \\
\midrule
SZZ & 0.49 & 0.60 & 0.54 & 0.46 & 0.61 & 0.53 & 0.48 & \textbf{0.68} & 0.57 & 0.41 & \textbf{0.58} & 0.48 \\
AG-SZZ & 0.48 & 0.58 & 0.52 & 0.46 & 0.64 & 0.53 & 0.45 & 0.59 & 0.51 & 0.31 & 0.42 & 0.36 \\
L-SZZ & 0.44 & 0.43 & 0.44 & 0.38 & 0.38 & 0.38 & 0.41 & 0.41 & 0.41 & 0.29 & 0.26 & 0.27 \\
R-SZZ & 0.48 & 0.47 & 0.47 & 0.59 & 0.58 & 0.59 & 0.47 & 0.47 & 0.47 & 0.40 & 0.36 & 0.38 \\
MA-SZZ & 0.45 & 0.57 & 0.50 & 0.45 & 0.64 & 0.53 & 0.43 & 0.56 & 0.49 & 0.30 & 0.42 & 0.35 \\
RA-SZZ & 0.45 & 0.57 & 0.50 & 0.45 & 0.64 & 0.53 & 0.37 & 0.48 & 0.42 & 0.24 & 0.36 & 0.29 \\
V-SZZ & 0.44 & 0.55 & 0.49 & 0.46 & 0.61 & 0.53 & 0.48 & 0.67 & 0.56 & 0.40 & 0.57 & 0.47 \\
LLM4SZZ & 0.65 & 0.64 & 0.64 & 0.63 & 0.62 & 0.63 & 0.55 & 0.55 & 0.55 & 0.56 & 0.48 & 0.51 \\
\midrule
\rowcolor{gray!20} \textbf{SZZ-Agent} & \textbf{0.78} & \textbf{0.76} & \textbf{0.77} & \textbf{0.75} & \textbf{0.74} & \textbf{0.75} & \textbf{0.67} & 0.67 & \textbf{0.67} & \textbf{0.65} & 0.56 & \textbf{0.60} \\
\bottomrule
\end{tabular}
\end{table*}

%% file: tables/table_data_leakage.tex
\begin{table}[t]
\footnotesize
\centering
\caption{RQ2a: SZZ-Agent vs.\ baselines on DS\_LINUX-26 (post training data cutoff). Best results in \textbf{bold}.}
\label{tab:results_leakage}
\begin{tabular}{lccc}
\toprule
\textbf{Approach} & \textbf{Precision} & \textbf{Recall} & \textbf{F1-Score} \\
\midrule
SZZ & 0.54 & 0.59 & 0.56 \\
AG-SZZ & 0.55 & 0.60 & 0.57 \\
L-SZZ & 0.52 & 0.51 & 0.51 \\
R-SZZ & 0.52 & 0.51 & 0.52 \\
MA-SZZ & 0.51 & 0.58 & 0.54 \\
RA-SZZ & 0.51 & 0.58 & 0.54 \\
V-SZZ & 0.44 & 0.49 & 0.47 \\
LLM4SZZ & 0.62 & 0.61 & 0.62 \\
\midrule
\rowcolor{gray!20} \textbf{SZZ-Agent (Ours)} & \textbf{0.81} & \textbf{0.80} & \textbf{0.81} \\
\bottomrule
\end{tabular}
\end{table}

%% file: tables/table_llm4szz_different_models.tex
\begin{table}[t]
\centering
\footnotesize
\caption{RQ2b: LLM4SZZ with different LLM backbones vs.\ SZZ-Agent. Best results in \textbf{bold}.}
\label{tab:results_models}
\begin{tabular}{lccc}
\toprule
\textbf{Approach} & \textbf{Precision} & \textbf{Recall} & \textbf{F1-Score} \\
\midrule
LLM4SZZ (gpt-4o-mini) & 0.65 & 0.64 & 0.64 \\
LLM4SZZ (gpt-5.2) & 0.61 & 0.60 & 0.61 \\
LLM4SZZ (claude-opus-4-5) & 0.66 & 0.65 & 0.65 \\
\midrule
\rowcolor{gray!20} \textbf{SZZ-Agent (claude-opus-4-5)} & \textbf{0.78} & \textbf{0.76} & \textbf{0.77} \\
\bottomrule
\end{tabular}
\end{table}

%% file: tables/table_without_message_or_diff.tex
\begin{table}[t]
\centering
\footnotesize
\caption{RQ2c: Impact of withholding the fix commit message and diff. Best results in \textbf{bold}.}
\label{tab:results_without_message_or_diff}
\begin{tabular}{lcc ccc}
\toprule
\textbf{Approach} & \makecell{\textbf{With}\\\textbf{Message}} & \makecell{\textbf{With}\\\textbf{Diff}} & \textbf{Precision} & \textbf{Recall} & \textbf{F1-Score} \\
\midrule
SZZ & -- & -- & 0.54 & 0.59 & 0.56 \\
SZZ-Agent & \ding{55} & \ding{55} & 0.49 & 0.48 & 0.48 \\
SZZ-Agent & \ding{55} & \checkmark & 0.74 & 0.73 & 0.73 \\
SZZ-Agent & \checkmark & \ding{55} & 0.78 & 0.77 & 0.77 \\
\midrule
\rowcolor{gray!20} \textbf{SZZ-Agent} & \checkmark & \checkmark & \textbf{0.81} & \textbf{0.80} & \textbf{0.81} \\
\bottomrule
\end{tabular}
\end{table}

%% file: tables/table_stage_01_vs_stage_02.tex
\begin{table}[t]
\footnotesize
\centering
\caption{RQ2d: Contribution of Stage 1 and Stage 2. Costs are average per fix commit. Best results in \textbf{bold}.}
\label{tab:stage_01_vs_stage_02}
\begin{tabular}{l ccc c}
\toprule
\textbf{Approach} & \textbf{Precision} & \textbf{Recall} & \textbf{F1-Score} & \textbf{Cost (USD)} \\
\midrule
Stage 1 Only & 0.48 & 0.47 & 0.48 & \textbf{\$0.18} \\
Stage 2 Only & 0.77 & 0.76 & 0.76 & \$1.20 \\
\midrule
\rowcolor{gray!20} \textbf{SZZ-Agent} & \textbf{0.81} & \textbf{0.80} & \textbf{0.81} & \$0.67 \\
\bottomrule
\end{tabular}
\end{table}

%% file: tables/table_commit_identifier_threshold.tex
\begin{table}[t]
\footnotesize
\centering
\caption{RQ2e: Impact of the commit selection threshold. Costs are average per fix commit. Best results in \textbf{bold}.}
\label{tab:commit_identifier_threshold}
\begin{tabular}{c ccc c}
\toprule
\makecell{\textbf{Commit} \\ \textbf{Identifier} \\ \textbf{Threshold}} & \textbf{Precision} & \textbf{Recall} & \textbf{F1-Score} & \textbf{Cost (USD)} \\
\midrule
8 & 0.76 & 0.75 & 0.76 & \$0.68 \\
32 & 0.81 & 0.80 & 0.81 & \$0.67 \\
128 & 0.80 & 0.79 & 0.80 & \$0.39 \\
512 & 0.81 & 0.80 & 0.81 & \$0.35 \\
$\infty$ & \textbf{0.82} & \textbf{0.81} & \textbf{0.82} & \textbf{\$0.33} \\
\bottomrule
\end{tabular}
\end{table}

%% file: sections/part_2.tex
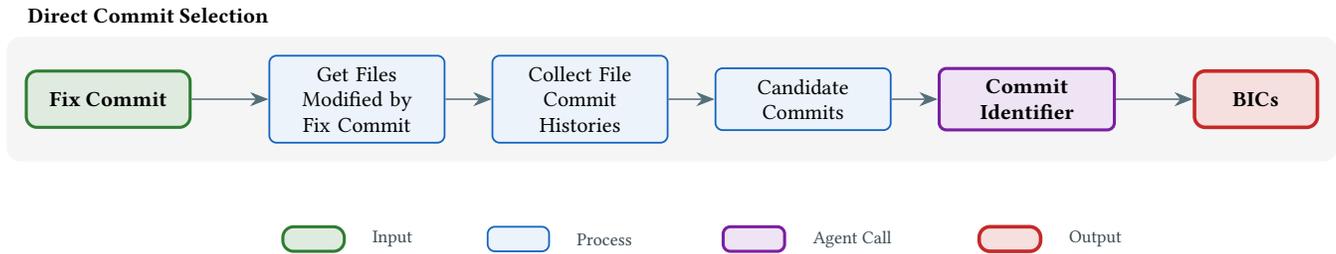
\begin{figure*}[t]
    \centering
    \resizebox{0.8\linewidth}{!}{\input{figures/simple_szz_agent_content}}
    \caption{Simple-SZZ-Agent: The fix commit's file histories are collected into a single candidate set. An agent with access to developer tools directly selects the bug-introducing commit, without SZZ-based filtering or binary search.}
    \label{fig:simple_szz_agent}
\end{figure*}

\input{tables/table_agent_comparison}

\section{Part 2: Simple-SZZ-Agent}
\label{sec:simple-szz-agent}
Our experiment on varying the commit selection threshold (RQ2e, \S\ref{sec:szz-agent:experiments:rq2e}) revealed that skipping binary search entirely and always using direct candidate selection yields the best performance at the lowest cost. If the agent can identify the bug-introducing commit from a large candidate set in a single step, do we need the complexity of SZZ-Agent at all? Motivated by this finding, we design Simple-SZZ-Agent, a drastically simplified approach that removes both the SZZ-based first stage and the binary search.

\subsection{Approach}
\label{sec:simple-szz-agent:approach}
Figure~\ref{fig:simple_szz_agent} shows an overview of Simple-SZZ-Agent. Given a fix commit, the approach collects the commit histories of all files modified by the fix, forming a single candidate set. It then provides an LLM-based agent with the fix commit message, the fix commit diff, and the full candidate set. The agent has access to standard developer tools (file reading, grep, glob) and is instructed to identify the bug-introducing commit. Since our ablations showed that omitting binary search improves both performance and cost (RQ2e), and that Stage~1 alone contributes less than Stage~2 (RQ2d), Simple-SZZ-Agent removes both: no SZZ-based filtering, no binary search, and no multi-stage pipeline---it relies entirely on the agent to find the bug-introducing commit among all candidates.

\begin{figure*}[t]
    \centering
    \begin{subfigure}[b]{0.48\textwidth}
        \centering
        \resizebox{0.9\textwidth}{!}{\input{figures/scatter_f1_vs_candidates}}
        \caption{F1-score.}
        \label{fig:scatter_f1_vs_candidates}
    \end{subfigure}\hfill
    \begin{subfigure}[b]{0.48\textwidth}
        \centering
        \resizebox{0.9\textwidth}{!}{\input{figures/scatter_cost_vs_candidates}}
        \caption{Cost per instance (USD).}
        \label{fig:scatter_cost_vs_candidates}
    \end{subfigure}

    \vspace{0.5em}

    \begin{subfigure}[b]{0.48\textwidth}
        \centering
        \resizebox{0.9\textwidth}{!}{\input{figures/scatter_tokens_vs_candidates}}
        \caption{Total tokens consumed.}
        \label{fig:scatter_tokens_vs_candidates}
    \end{subfigure}\hfill
    \begin{subfigure}[b]{0.48\textwidth}
        \centering
        \resizebox{0.9\textwidth}{!}{\input{figures/scatter_toolcalls_vs_candidates}}
        \caption{Total tool calls.}
        \label{fig:scatter_toolcalls_vs_candidates}
    \end{subfigure}
    \caption{Relationship between the number of candidate commits and F1-score, cost, token usage, and tool calls.}
    \label{fig:scatter_resource_vs_candidates}
\end{figure*}
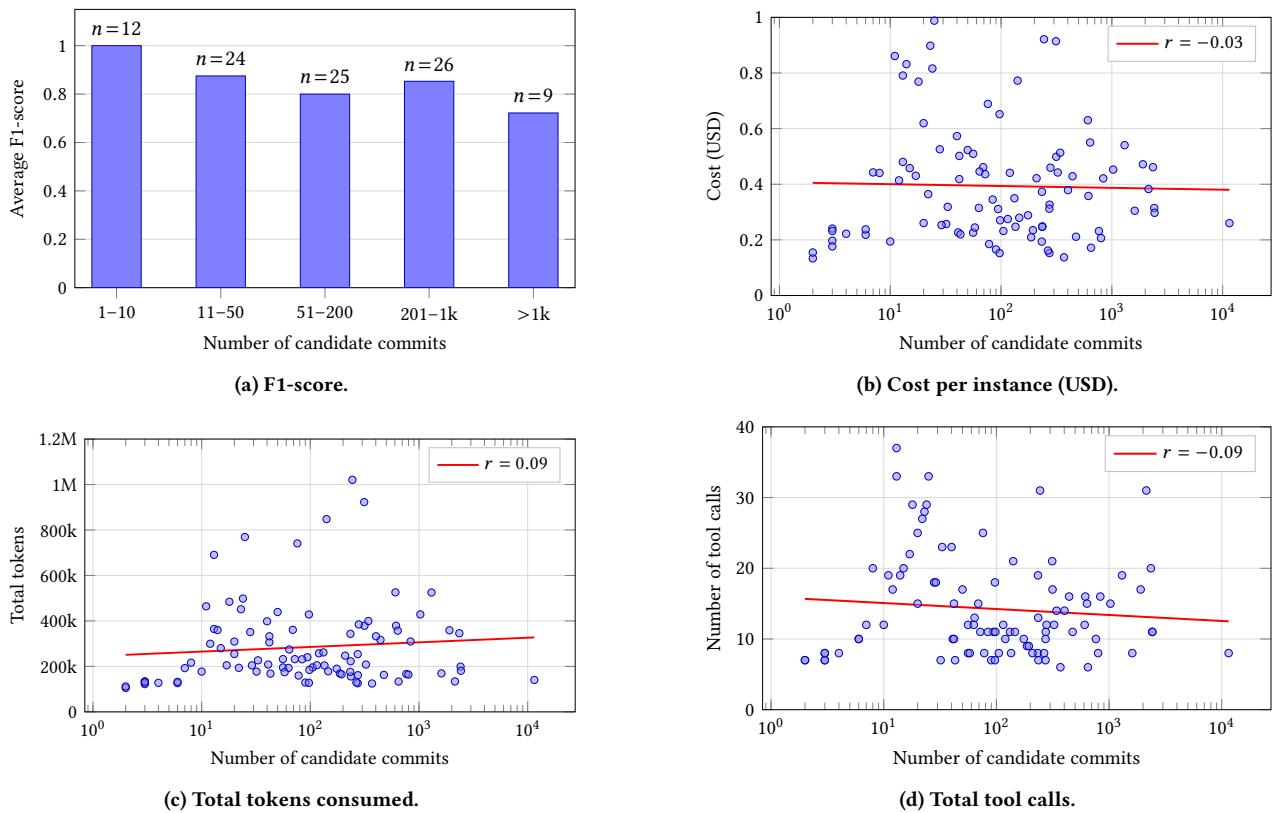

\subsection{Experimental Setup}
\label{sec:simple-szz-agent:experimental_setup}
We evaluate whether Simple-SZZ-Agent can match SZZ-Agent's performance and investigate why it works.

\begin{rqbox}
\textbf{RQ3} \textbf{(Effectiveness):} Can Simple-SZZ-Agent achieve a similar performance to SZZ-Agent at identifying bug-introducing commits?
\end{rqbox}

\begin{rqbox}
\textbf{RQ4} \textbf{(Analysis):} How can we explain the surprising success of Simple-SZZ-Agent?\\[2pt]
\hspace{1em}\textbf{(a) Strategy:} How can Simple-SZZ-Agent find the needle (BIC) in the haystack (large candidate commit set)?\\
\hspace{1em}\textbf{(b) Failures:} Where does Simple-SZZ-Agent still fail?\\
\hspace{1em}\textbf{(c) Generalization:} Can we expect even better results with better agents and/or better models?
\end{rqbox}

We use the same datasets, pre-processing, baselines, and metrics as in Part~1 (\S\ref{sec:szz-agent:experimental_setup}). Unless stated otherwise, all experiments use DS\_LINUX-26 with Claude Opus 4.5 as the LLM backbone and Claude Code as the agent framework.

\subsection{Experiments}
\label{sec:simple-szz-agent:experiments}

\subsubsection{RQ3: Effectiveness}
We evaluate whether Simple-SZZ-Agent can match or exceed the performance of SZZ-Agent despite its drastically simpler design.

\boldpar{Setup} We run Simple-SZZ-Agent on DS\_LINUX, DS\_GITHUB-c, DS\_GITHUB-j, and DS\_APACHE, and compare against SZZ, LLM4SZZ, and SZZ-Agent.

\boldpar{Results} Table~\ref{tab:agent_comparison} shows the results. Simple-SZZ-Agent matches or outperforms SZZ-Agent on every dataset. On DS\_LINUX, it achieves 0.81 F1-score compared to 0.77 for SZZ-Agent, an improvement of 4 percentage points. On DS\_GITHUB-c, both achieve 0.75. On DS\_GITHUB-j, Simple-SZZ-Agent achieves 0.75 compared to 0.67 for SZZ-Agent, an improvement of 8 percentage points. On DS\_APACHE, Simple-SZZ-Agent achieves 0.65 compared to 0.60 for SZZ-Agent, an improvement of 5 percentage points. Appendix~\ref{app:overlap} additionally reports the results of all approaches split by whether the fix commit and the bug-introducing commit overlap at the file and function level.

\result{\textbf{Finding 5:} Simple-SZZ-Agent matches or outperforms SZZ-Agent on all datasets, despite removing both the SZZ-based first stage and the binary search. A single agent call over the full candidate set is sufficient.}

\subsubsection{RQ4a: How Does the Agent Find the Needle in the Haystack?}
\label{sec:simple-szz-agent:experiments:rq4a}
Simple-SZZ-Agent receives hundreds or even thousands of candidate commits, yet it identifies the bug-introducing commit reliably. We investigate how.

\boldpar{Setup} We log all tool calls and analyze the behavior of the agent on DS\_LINUX-26 (100 fix commits). We record the relationship between the number of candidate commits and the agent's F1-score and resource consumption (cost, tokens, tool calls), the distribution of tool calls by type, and the sources of grep patterns.

\boldpar{Results} We expected that scanning a large candidate set would degrade performance and require substantially more resources. To our surprise, neither turned out to be the case. Figure~\ref{fig:scatter_f1_vs_candidates} shows the relationship between candidate set size and F1-score. Performance remains stable across candidate set sizes, with F1-scores above 0.80 for sets of up to 1,000 candidates. Only for the largest sets ($>$1,000 candidates) does F1-score decrease to 0.72. Figure~\ref{fig:scatter_resource_vs_candidates} shows the relationship between the number of candidate commits and resource consumption. Cost, token usage, and tool calls show no strong correlation with the number of candidates, indicating that the agent does not scale linearly with the candidate set size.

\begin{figure}[t]
    \centering
    \resizebox{\columnwidth}{!}{\input{figures/bar_toolcalls_per_tool}}
    \caption{Distribution of tool calls per tool type during agent execution.}
    \label{fig:bar_toolcalls_per_tool}
\end{figure}
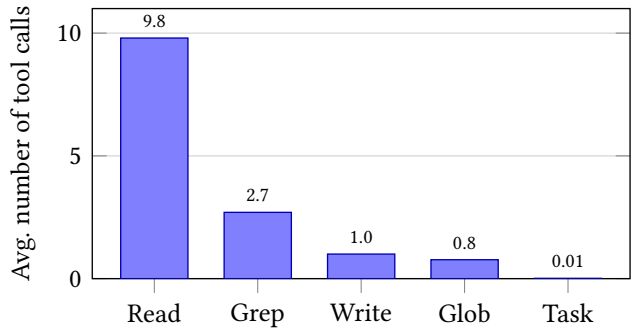

\begin{figure}[t]
    \centering
    \resizebox{\columnwidth}{!}{\input{figures/bar_grep_sources}}
    \caption{Distribution of grep search sources used by the agent. "FC" stands for "Fix Commit".}
    \label{fig:bar_grep_sources}
\end{figure}
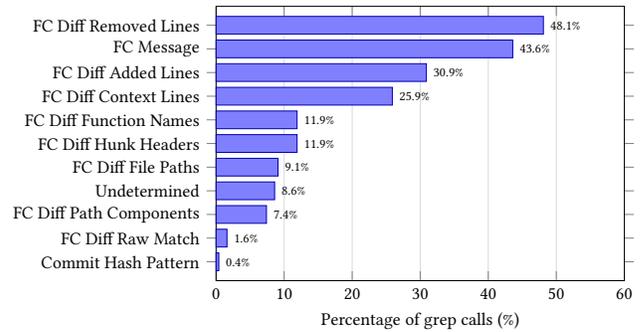

To understand how the agent achieves this, we examined its tool usage. Figure~\ref{fig:bar_toolcalls_per_tool} shows the distribution of tool calls. The agent predominantly uses Read (9.8 calls on average) and Grep (2.7 calls), with few calls to other tools. This reveals that the agent does not read all candidates sequentially; instead, it uses targeted searches to narrow down the candidate set. Figure~\ref{fig:bar_grep_sources} shows what the agent greps for. The most frequent grep patterns are derived from the removed lines of the fix commit diff (48.1\% of grep calls) and the fix commit message (43.6\%). The agent distills these sources into short patterns and uses them to search the candidate commits for matches. Across 243 grep calls, patterns have a median length of 21 characters (mean 25.2, std.\ dev.\ 14.8, min 6, max 104); the majority (63.8\%) are plain strings rather than regular expressions. 

Three examples illustrate this strategy:
\begin{enumerate}[leftmargin=*,nosep]
\item \textbf{Pattern from fix commit message.} The fix~\cite{linux_commit_1d66c3f2b8c0} replaces \texttt{fsleep()} with \texttt{udelay()} because \texttt{fsleep()} must not be called from atomic context. The agent greps for \texttt{"fsleep"} and directly locates the bug-introducing commit~\cite{linux_commit_01f60348d8fb}, which introduced the \texttt{fsleep()} call.
\item \textbf{Pattern from removed lines.} The fix~\cite{linux_commit_b4789aac9d34} corrects a firmware parser that assumed dword-aligned block sizes. The agent greps for \texttt{"blk->size >> 2"}, the faulty expression removed by the fix, and matches it to the bug-introducing commit~\cite{linux_commit_c6ed04f856a4} that added this line. This is analogous to what SZZ does---tracing removed lines back to the commit that introduced them---but the agent performs it via grep rather than \texttt{git blame}.
\item \textbf{Pure-addition fix.} The fix~\cite{linux_commit_ba1e9421cf1a} adds a NULL-pointer check for \texttt{smc\_ib\_is\_sg\_need\_sync()} (3 insertions, 0 deletions). Standard SZZ cannot handle this case at all, because there are no removed lines to trace back via \texttt{git blame}. The agent instead extracts the function name from the commit message, greps for \texttt{"smc\_ib\_is\_sg\_need\_sync"}, and identifies the bug-introducing commit~\cite{linux_commit_0ef69e788411} that introduced this function.
\end{enumerate}

\result{\textbf{Finding 6 (Key):} Cost and performance are largely independent of the candidate commit set size: the agent finds the needle without examining the full haystack. It does so by deriving short greppable patterns (median 21 characters) from the fix commit diff and message and using them to search the candidate set.}

\subsubsection{RQ4b: Where Does Simple-SZZ-Agent Fail?}
\label{sec:simple-szz-agent:experiments:rq4b}
We perform a manual error analysis to understand the failure cases.

\boldpar{Setup} We manually inspect all incorrect predictions of Simple-SZZ-Agent on DS\_LINUX-26 (100 fix commits). For each failure, we investigate the fix commit, the ground truth bug-introducing commit, the predicted bug-introducing commit, and the explanation provided by the agent, and categorize them by failure mode.

\boldpar{Results} Figure~\ref{fig:sankey_error_analysis} shows the results. Of 100 fix commits, Simple-SZZ-Agent correctly identifies the bug-introducing commit in 87 cases. Of the 13 incorrect predictions, 4 fail because the true bug-introducing commit is not in the candidate set (i.e., the bug was introduced in a file not modified by the fix). The remaining 9 failures occur when the bug-introducing commit is in the candidate set but the agent does not select the correct one. 

We further categorize these 9 cases through manual analysis: in 3 cases, the agent performs an incorrect analysis---it identifies the correct code area but confuses commits that modified the relevant code with the commit that originally introduced the flaw; in 2 cases, the agent selects a commit within 3 positions of the correct one (near miss), including one case where the predicted and ground truth commits are consecutive commits from the same patch series; in 3 cases, the ground truth label itself is questionable; and in 1 case, we could not determine the reason for the failure because the fix and ground truth bug-introducing commits were too complex for us to analyze as non-Linux-kernel experts. Since we are not Linux kernel experts, we also cannot determine with certainty whether the 3 questionable labels are incorrect; we include a detailed report for each case in our artifact.

\result{\textbf{Finding 7:} Of the 13 failures, 4 occur because the bug-introducing commit is not in the candidate set. Of the remaining 9, 3 are due to incorrect agent analysis, 2 are near misses, 3 involve questionable ground truth labels, and 1 we could not categorize.}

\subsubsection{RQ4c: Better Models and Agents}
\label{sec:simple-szz-agent:experiments:rq4c}
We investigate whether Simple-SZZ-Agent's performance generalizes across different LLM backbones and agent frameworks, and whether cheaper or more capable configurations can improve results.

\boldpar{Setup} We evaluate Simple-SZZ-Agent with two agent frameworks---Claude Code and OpenHands---across multiple LLM backbones: Claude Haiku 4.5, Claude Sonnet 4.5, Claude Opus 4.5, as well as minimax-m2.5 and glm-5 for OpenHands. All experiments use DS\_LINUX-26 (100 samples).

\boldpar{Results} Table~\ref{tab:results_cheaper_models} shows the results. Simple-SZZ-Agent is effective across all tested configurations. Even the cheapest configuration (OpenHands with minimax-m2.5 at \$0.04 per fix commit) achieves 0.79 F1-score, outperforming all previous non-agentic approaches. The best configuration (OpenHands with Claude Sonnet 4.5) achieves 0.86 F1-score at \$0.38 per fix commit, surpassing the Claude Opus 4.5 configurations. Within the same agent framework, more capable models generally perform better, though the difference between Sonnet and Opus is small or even reversed. The OpenHands configurations consume three to four times more tokens than their Claude Code counterparts, so their higher scores may be attributable to higher token usage rather than architectural differences. Equalizing token budgets across setups is not straightforward: OpenHands only supports hard cutoffs (maximum budget or iterations) that terminate execution, and Claude Opus 4.5 lacks the fine-grained thinking-effort controls of later models. Appendix~\ref{app:effort-split} breaks down the effort of each configuration by prediction outcome: in almost all configurations, unsuccessful predictions consume more resources than successful ones.

\result{\textbf{Finding 8:} Simple-SZZ-Agent generalizes across agent frameworks and LLM backbones. Even the worst-performing configuration (Claude Code with Claude Haiku 4.5, 0.74 F1) outperforms the previous state of the art by 12 percentage points, and the best configuration reaches 0.86 F1-score, suggesting that further progress with improved models and agents is likely.}

%% file: figures/simple_szz_agent_content.tex
\begin{tikzpicture}[
    node distance=0.4cm and 0.5cm,
    >={Stealth[length=1.8mm, width=1.4mm]},
    inputnode/.style={
        rectangle, rounded corners=3pt,
        minimum width=1.6cm, minimum height=0.55cm,
        text centered, draw=inputgreen, line width=0.85pt,
        fill=inputgreen!15, font=\scriptsize\bfseries
    },
    procnode/.style={
        rectangle, rounded corners=2pt,
        minimum width=1.6cm, minimum height=0.55cm,
        text centered, draw=processblue, line width=0.5pt,
        fill=processblue!8, font=\scriptsize,
        text width=1.5cm, align=center
    },
    agentnode/.style={
        rectangle, rounded corners=2pt,
        minimum width=1.6cm, minimum height=0.55cm,
        text centered, draw=agentpurple, line width=0.85pt,
        fill=agentpurple!12, font=\scriptsize\bfseries,
        text width=1.5cm, align=center
    },
    outnode/.style={
        rectangle, rounded corners=3pt,
        minimum width=1.2cm, minimum height=0.55cm,
        text centered, draw=outputred, line width=1pt,
        fill=outputred!15, font=\scriptsize\bfseries
    },
    myarrow/.style={->, line width=0.42pt, draw=arrowgray}
]


\node[inputnode] (start) {Fix Commit};

\node[procnode, right=0.75cm of start] (getfiles) {Get Files\\Modified by\\Fix Commit};

\node[procnode, right=0.45cm of getfiles] (gethist) {Collect File\\Commit Histories};

\node[procnode, right=0.45cm of gethist] (cand) {Candidate\\Commits};

\node[agentnode, right=0.45cm of cand] (cia) {Commit\\Identifier};

\node[outnode, right=0.75cm of cia] (out) {BICs};


\begin{scope}[on background layer]
    \node[fit=(start)(getfiles)(gethist)(cand)(cia)(out),
          fill=stage1bg, rounded corners=4pt, inner sep=5pt] (bg) {};
\end{scope}


\node[font=\scriptsize\bfseries, text=black, anchor=north west]
    at ($(bg.north west) + (0.1, +0.4)$) {Direct Commit Selection};


\draw[myarrow] (start) -- (getfiles);
\draw[myarrow] (getfiles) -- (gethist);
\draw[myarrow] (gethist) -- (cand);
\draw[myarrow] (cand) -- (cia);
\draw[myarrow] (cia) -- (out);


\begin{scope}[shift={($(bg.south) + (0, -0.75)$)}]
    \node[inputnode, minimum width=0.6cm, minimum height=0.24cm, font=\tiny\bfseries, inner sep=0.3pt] at (-3.5,0) (L1) {};
    \node[font=\tiny, text=textgray, anchor=west] at ($(L1.east)+(0.15,0)$) {Input};

    \node[procnode, minimum width=0.6cm, minimum height=0.24cm, font=\tiny, inner sep=0.3pt, text width=0.4cm] at (-1.5,0) (L2) {};
    \node[font=\tiny, text=textgray, anchor=west] at ($(L2.east)+(0.15,0)$) {Process};

    \node[agentnode, minimum width=0.6cm, minimum height=0.24cm, font=\tiny\bfseries, inner sep=0.3pt, text width=0.4cm] at (0.8,0) (L3) {};
    \node[font=\tiny, text=textgray, anchor=west] at ($(L3.east)+(0.15,0)$) {Agent Call};

    \node[outnode, minimum width=0.6cm, minimum height=0.24cm, font=\tiny\bfseries, inner sep=0.3pt] at (3.3,0) (L5) {};
    \node[font=\tiny, text=textgray, anchor=west] at ($(L5.east)+(0.15,0)$) {Output};
\end{scope}

\end{tikzpicture}

%% file: tables/table_agent_comparison.tex
\begin{table*}[t]
\footnotesize
\centering
\caption{RQ3: Simple-SZZ-Agent vs.\ SZZ-Agent across four datasets. Best results per dataset in \textbf{bold}.}
\label{tab:agent_comparison}
\begin{tabular}{lcccccccccccc}
\toprule
 & \multicolumn{3}{c}{\textbf{DS\_LINUX}} & \multicolumn{3}{c}{\textbf{DS\_GITHUB-c}} & \multicolumn{3}{c}{\textbf{DS\_GITHUB-j}} & \multicolumn{3}{c}{\textbf{DS\_APACHE}} \\
\cmidrule(lr){2-4} \cmidrule(lr){5-7} \cmidrule(lr){8-10} \cmidrule(lr){11-13}
\textbf{Approach} & \textbf{Precision} & \textbf{Recall} & \textbf{F1-Score} & \textbf{Precision} & \textbf{Recall} & \textbf{F1-Score} & \textbf{Precision} & \textbf{Recall} & \textbf{F1-Score} & \textbf{Precision} & \textbf{Recall} & \textbf{F1-Score} \\
\midrule
SZZ & 0.49 & 0.60 & 0.54 & 0.46 & 0.61 & 0.53 & 0.48 & 0.68 & 0.57 & 0.41 & 0.58 & 0.48 \\
LLM4SZZ & 0.65 & 0.64 & 0.64 & 0.63 & 0.62 & 0.63 & 0.55 & 0.55 & 0.55 & 0.56 & 0.48 & 0.51 \\
SZZ-Agent & 0.78 & 0.76 & 0.77 & \textbf{0.75} & 0.74 & \textbf{0.75} & 0.67 & 0.67 & 0.67 & 0.65 & 0.56 & 0.60 \\
\midrule
\rowcolor{gray!20} \textbf{Simple-SZZ-Agent} & \textbf{0.82} & \textbf{0.81} & \textbf{0.81} & \textbf{0.75} & \textbf{0.75} & \textbf{0.75} & \textbf{0.75} & \textbf{0.75} & \textbf{0.75} & \textbf{0.70} & \textbf{0.60} & \textbf{0.65} \\
\bottomrule
\end{tabular}
\end{table*}

%% file: figures/scatter_f1_vs_candidates.tex
\begin{tikzpicture}
\begin{axis}[
    width=1.02\columnwidth,
    height=0.56\columnwidth,
    ybar,
    bar width=20pt,
    xlabel={Number of candidate commits},
    ylabel={Average F1-score},
    ymin=0, ymax=1.15,
    ytick={0, 0.2, 0.4, 0.6, 0.8, 1.0},
    symbolic x coords={1--10,11--50,51--200,201--1k,{$>$1k}},
    xtick=data,
    x tick label style={font=\small},
    tick label style={font=\small},
    label style={font=\small},
    grid=major,
    grid style={gray!30},
    ymajorgrids=true,
    xmajorgrids=false,
    nodes near coords={\scriptsize $n\!=\!{}$},
    every node near coord/.append style={anchor=south, yshift=1pt},
    xtick pos=left,
]
\addplot[
    fill=blue!50,
    draw=blue!70!black,
    point meta=explicit symbolic,
    nodes near coords={\pgfplotspointmeta},
] coordinates {
    (1--10, 1.000) [$n\!=\!12$]
    (11--50, 0.875) [$n\!=\!24$]
    (51--200, 0.800) [$n\!=\!25$]
    (201--1k, 0.853) [$n\!=\!26$]
    ({$>$1k}, 0.722) [$n\!=\!9$]
};
\end{axis}
\end{tikzpicture}

%% file: figures/scatter_cost_vs_candidates.tex
\begin{tikzpicture}
\begin{axis}[
    width=1.02\columnwidth,
    height=0.56\columnwidth,
    xlabel={Number of candidate commits},
    ylabel={Cost (USD)},
    xmode=log,
    ymin=0, ymax=1.0,
    grid=major,
    grid style={gray!30},
    tick label style={font=\small},
    label style={font=\small},
    legend style={at={(0.97,0.97)}, anchor=north east, font=\small, draw=gray!50},
]
\addplot[
    only marks,
    mark=*,
    mark size=1.5pt,
    draw=blue!70!black,
    fill=blue!50,
    fill opacity=0.5,
    forget plot,
] coordinates {
    (42, 0.418451)
    (475, 0.211396)
    (274, 0.326675)
    (7, 0.442582)
    (175, 0.288381)
    (244, 0.921255)
    (90, 0.165735)
    (370, 0.137343)
    (2412, 0.314327)
    (236, 0.248515)
    (11509, 0.260068)
    (69, 0.461320)
    (23, 0.898165)
    (764, 0.231893)
    (799, 0.206413)
    (32, 0.257172)
    (120, 0.441195)
    (273, 0.152673)
    (234, 0.372971)
    (2, 0.133588)
    (56, 0.226088)
    (12, 0.413906)
    (273, 0.312164)
    (28, 0.525757)
    (6, 0.218423)
    (442, 0.428757)
    (315, 0.498876)
    (50, 0.522966)
    (64, 0.445895)
    (10, 0.194079)
    (18, 0.769010)
    (233, 0.193941)
    (2141, 0.382842)
    (40, 0.573201)
    (141, 0.772663)
    (97, 0.651778)
    (72, 0.436165)
    (105, 0.231778)
    (615, 0.357999)
    (78, 0.184980)
    (135, 0.247289)
    (41, 0.226874)
    (20, 0.619478)
    (1907, 0.471841)
    (4, 0.222088)
    (20, 0.260423)
    (43, 0.219832)
    (313, 0.914072)
    (115, 0.274958)
    (132, 0.349562)
    (76, 0.688845)
    (6, 0.237742)
    (2429, 0.297378)
    (3, 0.197118)
    (8, 0.440652)
    (22, 0.364396)
    (2, 0.154645)
    (187, 0.209337)
    (1028, 0.452809)
    (13, 0.480222)
    (3, 0.240752)
    (14, 0.831731)
    (3, 0.176451)
    (235, 0.246750)
    (649, 0.171775)
    (3, 0.232135)
    (94, 0.311084)
    (146, 0.279483)
    (2350, 0.461356)
    (56, 0.509436)
    (24, 0.816198)
    (11, 0.861147)
    (836, 0.421093)
    (1608, 0.304719)
    (342, 0.513304)
    (1303, 0.540460)
    (84, 0.345060)
    (42, 0.501950)
    (29, 0.253306)
    (98, 0.270364)
    (97, 0.152357)
    (265, 0.161554)
    (325, 0.442334)
    (17, 0.430683)
    (402, 0.378843)
    (607, 0.630233)
    (33, 0.318761)
    (15, 0.457899)
    (25, 0.988192)
    (194, 0.234865)
    (63, 0.315023)
    (58, 0.244535)
    (279, 0.459615)
    (637, 0.550160)
    (209, 0.421615)
    (13, 0.790883)
};
\addplot[
    red, thick, no markers,
] coordinates {
    (2.0000, 0.404921)
    (2.3865, 0.404413)
    (2.8477, 0.403906)
    (3.3981, 0.403399)
    (4.0548, 0.402891)
    (4.8384, 0.402384)
    (5.7735, 0.401877)
    (6.8893, 0.401369)
    (8.2207, 0.400862)
    (9.8094, 0.400355)
    (11.7052, 0.399847)
    (13.9673, 0.399340)
    (16.6666, 0.398833)
    (19.8876, 0.398325)
    (23.7311, 0.397818)
    (28.3174, 0.397311)
    (33.7900, 0.396803)
    (40.3202, 0.396296)
    (48.1124, 0.395789)
    (57.4106, 0.395281)
    (68.5057, 0.394774)
    (81.7451, 0.394267)
    (97.5431, 0.393759)
    (116.3942, 0.393252)
    (138.8885, 0.392745)
    (165.7300, 0.392237)
    (197.7589, 0.391730)
    (235.9777, 0.391223)
    (281.5826, 0.390715)
    (336.0010, 0.390208)
    (400.9364, 0.389701)
    (478.4211, 0.389193)
    (570.8804, 0.388686)
    (681.2084, 0.388179)
    (812.8583, 0.387671)
    (969.9508, 0.387164)
    (1157.4029, 0.386657)
    (1381.0818, 0.386149)
    (1647.9888, 0.385642)
    (1966.4780, 0.385135)
    (2346.5183, 0.384627)
    (2800.0050, 0.384120)
    (3341.1321, 0.383613)
    (3986.8372, 0.383105)
    (4757.3307, 0.382598)
    (5676.7294, 0.382091)
    (6773.8105, 0.381583)
    (8082.9128, 0.381076)
    (9645.0113, 0.380569)
    (11509.0000, 0.380061)
};
\addlegendentry{$r = -0.03$}
\end{axis}
\end{tikzpicture}

%% file: figures/scatter_tokens_vs_candidates.tex
\begin{tikzpicture}
\begin{axis}[
    width=1.02\columnwidth,
    height=0.56\columnwidth,
    xlabel={Number of candidate commits},
    ylabel={Total tokens},
    xmode=log,
    ymin=0, ymax=1200000,
    ytick={0,200000,400000,600000,800000,1000000,1200000},
    yticklabels={0,200k,400k,600k,800k,1M,1.2M},
    scaled y ticks=false,
    grid=major,
    grid style={gray!30},
    tick label style={font=\small},
    label style={font=\small},
    legend style={at={(0.97,0.97)}, anchor=north east, font=\small, draw=gray!50},
]
\addplot[
    only marks,
    mark=*,
    mark size=1.5pt,
    draw=blue!70!black,
    fill=blue!50,
    fill opacity=0.5,
    forget plot,
] coordinates {
    (42, 306009)
    (475, 162743)
    (274, 161353)
    (7, 192512)
    (175, 189780)
    (244, 1020754)
    (90, 128954)
    (370, 124783)
    (2412, 199088)
    (236, 155863)
    (11509, 140470)
    (69, 360877)
    (23, 451399)
    (764, 166503)
    (799, 164095)
    (32, 177051)
    (120, 257655)
    (273, 125949)
    (234, 343197)
    (2, 105447)
    (56, 195198)
    (12, 299711)
    (273, 253937)
    (28, 350978)
    (6, 126509)
    (442, 316561)
    (315, 378724)
    (50, 439201)
    (64, 274425)
    (10, 177125)
    (18, 484344)
    (233, 176076)
    (2141, 133814)
    (40, 398437)
    (141, 847809)
    (97, 428321)
    (72, 231913)
    (105, 195642)
    (615, 379075)
    (78, 159724)
    (135, 204105)
    (41, 208293)
    (20, 309336)
    (1907, 358670)
    (4, 127796)
    (20, 254584)
    (43, 168226)
    (313, 922612)
    (115, 205088)
    (132, 261501)
    (76, 741034)
    (6, 132223)
    (2429, 180755)
    (3, 122219)
    (8, 216501)
    (22, 193844)
    (2, 111939)
    (187, 168770)
    (1028, 428668)
    (13, 364449)
    (3, 133447)
    (14, 359987)
    (3, 133934)
    (235, 222600)
    (649, 133477)
    (3, 129088)
    (94, 240197)
    (146, 178252)
    (2350, 345809)
    (56, 232075)
    (24, 498391)
    (11, 464285)
    (836, 309511)
    (1608, 169127)
    (342, 399411)
    (1303, 525055)
    (84, 231286)
    (42, 332364)
    (29, 204563)
    (98, 184661)
    (97, 127892)
    (265, 129471)
    (325, 208112)
    (17, 204913)
    (402, 332453)
    (607, 525721)
    (33, 226690)
    (15, 280268)
    (25, 769169)
    (194, 165345)
    (63, 193336)
    (58, 174425)
    (279, 384619)
    (637, 357175)
    (209, 247264)
    (13, 690640)
};
\addplot[
    red, thick, no markers,
] coordinates {
    (2.0000, 250994.4)
    (2.3865, 252553.4)
    (2.8477, 254112.4)
    (3.3981, 255671.4)
    (4.0548, 257230.3)
    (4.8384, 258789.3)
    (5.7735, 260348.3)
    (6.8893, 261907.3)
    (8.2207, 263466.3)
    (9.8094, 265025.2)
    (11.7052, 266584.2)
    (13.9673, 268143.2)
    (16.6666, 269702.2)
    (19.8876, 271261.2)
    (23.7311, 272820.1)
    (28.3174, 274379.1)
    (33.7900, 275938.1)
    (40.3202, 277497.1)
    (48.1124, 279056.1)
    (57.4106, 280615.0)
    (68.5057, 282174.0)
    (81.7451, 283733.0)
    (97.5431, 285292.0)
    (116.3942, 286850.9)
    (138.8885, 288409.9)
    (165.7300, 289968.9)
    (197.7589, 291527.9)
    (235.9777, 293086.9)
    (281.5826, 294645.8)
    (336.0010, 296204.8)
    (400.9364, 297763.8)
    (478.4211, 299322.8)
    (570.8804, 300881.8)
    (681.2084, 302440.7)
    (812.8583, 303999.7)
    (969.9508, 305558.7)
    (1157.4029, 307117.7)
    (1381.0818, 308676.7)
    (1647.9888, 310235.6)
    (1966.4780, 311794.6)
    (2346.5183, 313353.6)
    (2800.0050, 314912.6)
    (3341.1321, 316471.6)
    (3986.8372, 318030.5)
    (4757.3307, 319589.5)
    (5676.7294, 321148.5)
    (6773.8105, 322707.5)
    (8082.9128, 324266.5)
    (9645.0113, 325825.4)
    (11509.0000, 327384.4)
};
\addlegendentry{$r = 0.09$}
\end{axis}
\end{tikzpicture}

%% file: figures/scatter_toolcalls_vs_candidates.tex
\begin{tikzpicture}
\begin{axis}[
    width=1.02\columnwidth,
    height=0.56\columnwidth,
    xlabel={Number of candidate commits},
    ylabel={Number of tool calls},
    xmode=log,
    ymin=0, ymax=40,
    grid=major,
    grid style={gray!30},
    tick label style={font=\small},
    label style={font=\small},
    legend style={at={(0.97,0.97)}, anchor=north east, font=\small, draw=gray!50},
]
\addplot[
    only marks,
    mark=*,
    mark size=1.5pt,
    draw=blue!70!black,
    fill=blue!50,
    fill opacity=0.5,
    forget plot,
] coordinates {
    (42, 15)
    (475, 11)
    (274, 11)
    (7, 12)
    (175, 10)
    (244, 31)
    (90, 7)
    (370, 6)
    (2412, 11)
    (236, 7)
    (11509, 8)
    (69, 15)
    (23, 28)
    (764, 10)
    (799, 8)
    (32, 7)
    (120, 10)
    (273, 7)
    (234, 19)
    (2, 7)
    (56, 8)
    (12, 17)
    (273, 10)
    (28, 18)
    (6, 10)
    (442, 16)
    (315, 17)
    (50, 17)
    (64, 13)
    (10, 12)
    (18, 29)
    (233, 8)
    (2141, 31)
    (40, 23)
    (141, 21)
    (97, 18)
    (72, 11)
    (105, 8)
    (615, 16)
    (78, 8)
    (135, 8)
    (41, 10)
    (20, 25)
    (1907, 17)
    (4, 8)
    (20, 15)
    (43, 7)
    (313, 21)
    (115, 12)
    (132, 11)
    (76, 25)
    (6, 10)
    (2429, 11)
    (3, 7)
    (8, 20)
    (22, 27)
    (2, 7)
    (187, 9)
    (1028, 15)
    (13, 33)
    (3, 7)
    (14, 19)
    (3, 8)
    (235, 13)
    (649, 6)
    (3, 8)
    (94, 11)
    (146, 11)
    (2350, 20)
    (56, 12)
    (24, 29)
    (11, 19)
    (836, 16)
    (1608, 8)
    (342, 14)
    (1303, 19)
    (84, 11)
    (42, 10)
    (29, 18)
    (98, 11)
    (97, 7)
    (265, 8)
    (325, 12)
    (17, 22)
    (402, 14)
    (607, 12)
    (33, 23)
    (15, 20)
    (25, 33)
    (194, 9)
    (63, 12)
    (58, 8)
    (279, 12)
    (637, 15)
    (209, 8)
    (13, 37)
};
\addplot[
    red, thick, no markers,
] coordinates {
    (2.0000, 15.6732)
    (2.3865, 15.6084)
    (2.8477, 15.5435)
    (3.3981, 15.4786)
    (4.0548, 15.4138)
    (4.8384, 15.3489)
    (5.7735, 15.2841)
    (6.8893, 15.2192)
    (8.2207, 15.1544)
    (9.8094, 15.0895)
    (11.7052, 15.0247)
    (13.9673, 14.9598)
    (16.6666, 14.8950)
    (19.8876, 14.8301)
    (23.7311, 14.7652)
    (28.3174, 14.7004)
    (33.7900, 14.6355)
    (40.3202, 14.5707)
    (48.1124, 14.5058)
    (57.4106, 14.4410)
    (68.5057, 14.3761)
    (81.7451, 14.3113)
    (97.5431, 14.2464)
    (116.3942, 14.1816)
    (138.8885, 14.1167)
    (165.7300, 14.0518)
    (197.7589, 13.9870)
    (235.9777, 13.9221)
    (281.5826, 13.8573)
    (336.0010, 13.7924)
    (400.9364, 13.7276)
    (478.4211, 13.6627)
    (570.8804, 13.5979)
    (681.2084, 13.5330)
    (812.8583, 13.4682)
    (969.9508, 13.4033)
    (1157.4029, 13.3384)
    (1381.0818, 13.2736)
    (1647.9888, 13.2087)
    (1966.4780, 13.1439)
    (2346.5183, 13.0790)
    (2800.0050, 13.0142)
    (3341.1321, 12.9493)
    (3986.8372, 12.8845)
    (4757.3307, 12.8196)
    (5676.7294, 12.7548)
    (6773.8105, 12.6899)
    (8082.9128, 12.6251)
    (9645.0113, 12.5602)
    (11509.0000, 12.4953)
};
\addlegendentry{$r = -0.09$}
\end{axis}
\end{tikzpicture}

%% file: figures/bar_toolcalls_per_tool.tex
\begin{tikzpicture}
\begin{axis}[
    width=0.85\columnwidth,
    height=0.51\columnwidth,
    ybar,
    bar width=20pt,
    ylabel={Avg.\ number of tool calls},
    ymin=0,
    ymax=11,
    symbolic x coords={Read,Grep,Write,Glob,Task},
    xtick=data,
    x tick label style={font=\small},
    tick label style={font=\small},
    label style={font=\small},
    grid=major,
    grid style={gray!30},
    ymajorgrids=true,
    xmajorgrids=false,
    point meta=explicit symbolic,
    nodes near coords={\pgfplotspointmeta},
    nodes near coords style={font=\scriptsize, anchor=south},
    enlarge x limits=0.15,
    xtick pos=left,
]
\addplot[
    fill=blue!50,
    draw=blue!70!black,
] coordinates {
    (Read, 9.80) [9.8]
    (Grep, 2.70) [2.7]
    (Write, 1.00) [1.0]
    (Glob, 0.77) [0.8]
    (Task, 0.01) [0.01]
};
\end{axis}
\end{tikzpicture}

%% file: figures/bar_grep_sources.tex
\begin{tikzpicture}
\begin{axis}[
    width=0.95\columnwidth,
    height=0.95\columnwidth,
    xbar,
    bar width=11pt,
    xlabel={Percentage of grep calls (\%)},
    xmin=0,
    xmax=60,
    symbolic y coords={{Commit Hash Pattern},{FC Diff Raw Match},{FC Diff Path Components},{Undetermined},{FC Diff File Paths},{FC Diff Hunk Headers},{FC Diff Function Names},{FC Diff Context Lines},{FC Diff Added Lines},{FC Message},{FC Diff Removed Lines}},
    ytick=data,
    ytick pos=left,
    y tick label style={font=\small, anchor=east},
    tick label style={font=\small},
    label style={font=\small},
    grid=major,
    grid style={gray!30},
    xmajorgrids=true,
    ymajorgrids=false,
    point meta=explicit symbolic,
    nodes near coords={\pgfplotspointmeta},
    nodes near coords style={font=\scriptsize, anchor=west},
    enlarge y limits=0.08,
    clip=false,
]
\addplot[
    fill=blue!50,
    draw=blue!70!black,
] coordinates {
    (0.4,{Commit Hash Pattern}) [0.4\%]
    (1.6,{FC Diff Raw Match}) [1.6\%]
    (7.4,{FC Diff Path Components}) [7.4\%]
    (8.6,{Undetermined}) [8.6\%]
    (9.1,{FC Diff File Paths}) [9.1\%]
    (11.9,{FC Diff Hunk Headers}) [11.9\%]
    (11.9,{FC Diff Function Names}) [11.9\%]
    (25.9,{FC Diff Context Lines}) [25.9\%]
    (30.9,{FC Diff Added Lines}) [30.9\%]
    (43.6,{FC Message}) [43.6\%]
    (48.1,{FC Diff Removed Lines}) [48.1\%]
};
\end{axis}
\end{tikzpicture}

%% file: sections/discussion.tex
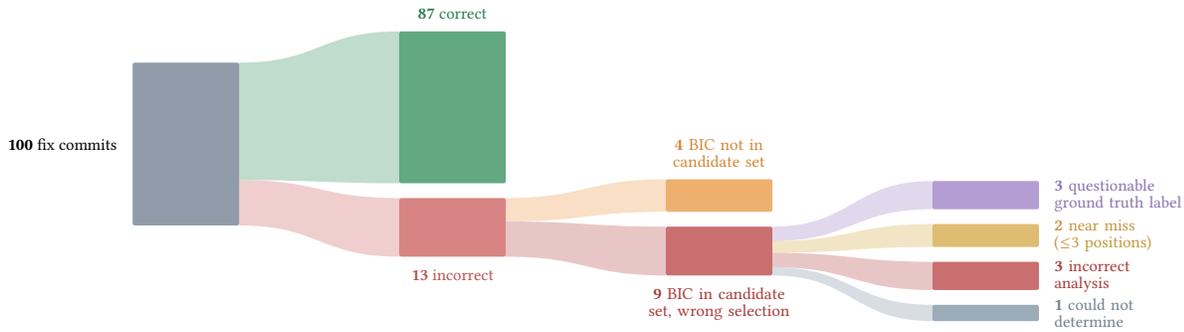
\begin{figure*}[t]
    \centering
    \resizebox{0.95\textwidth}{!}{\input{figures/sankey_error_analysis}}
    \caption{Manual error analysis of Simple-SZZ-Agent on 100 fix commits from DS\_LINUX-26.}
    \label{fig:sankey_error_analysis}
\end{figure*}

\input{tables/table_cheaper_models}

\newpage\section{Discussion}
\label{sec:discussion}
We discuss directions for future work motivated by our findings. While these directions are beyond the scope of this paper, we hope they inspire other researchers to explore them. We cover research on identifying bug-introducing commits (SZZ research), root cause understanding, and program repair.

\subsection{SZZ Research}

\boldpar{Closing the Gap} Our results represent a substantial step toward solving the problem of identifying bug-introducing commits from fix commits. Simple-SZZ-Agent achieves up to 0.86 F1-score on the Linux kernel dataset, compared to 0.62 for the previous state of the art. Still, the problem is not fully solved. Our error analysis~(\S\ref{sec:simple-szz-agent:experiments:rq4b}) revealed that of the 13 failures, 4 occur because the bug-introducing commit is not in the candidate set, 3 are due to incorrect agent analysis, 2 are near misses, 3 involve questionable ground truth labels, and 1 we could not categorize. These failure modes point to two main areas for improvement.

First, the incorrect analyses and near misses (5 cases total) suggest that the agent's reasoning about which commit introduced a bug can still be improved. Our results in~\S\ref{sec:simple-szz-agent:experiments:rq4c} show that different model and agent combinations yield meaningful performance differences, indicating that continued model improvements are a promising avenue.

Second, the 4 cases where the bug-introducing commit is not in the candidate set reveal a limitation of the current candidate collection strategy: Simple-SZZ-Agent only searches the commit histories of files modified by the fix, so when the bug was introduced in a different file, the correct commit is never considered. Appendix~\ref{app:overlap} quantifies this limitation across all datasets: it affects 5.5\% to 13.2\% of cases, and the baselines also fail almost completely on this subset. How to expand the candidate set beyond the files touched by the fix without introducing prohibitive infrastructure costs is an open question that we leave for future work.

\subsection{Root Cause Understanding}

Our analysis in~\S\ref{sec:simple-szz-agent:experiments:rq4a} revealed that agents compress information about a bug into short greppable patterns that locate its introducing changes. This compression itself may have value beyond identifying the bug-introducing commit: it captures, in a concise form, what made the code buggy.

A practitioner who knows how to fix a bug does not always understand how it was introduced. The patterns the agent derives and the bug-introducing changes it identifies could help provide this understanding. Our context ablation~(\S\ref{sec:szz-agent:experiments:rq2c}) showed that the agent achieves 0.73 F1-score with only the diff and no commit message, demonstrating that it can extract the essential characteristics of a bug from code changes alone. Whether these patterns can be used to generate useful explanations of how a bug was introduced---especially when the fix commit message is uninformative---is an open question for future work.

\subsection{Program Repair}

In program repair, the goal is to generate fixes for bugs. Our approach currently operates after a fix has already been written: it takes the fix commit as input and traces back to the bug-introducing commit. An open question is whether the bug-introducing commit can also be identified from the outputs of bug oracles instead of fix commits---for example, from crash reports of a fuzzer, failing unit tests, or alerts from monitoring software. If so, agents could provide the bug-introducing changes as context for generating the fix itself, enabling agents to both locate and fix bugs. We leave this question for future work.

%% file: figures/sankey_error_analysis.tex
\begin{tikzpicture}[x=1cm, y=1cm]

\definecolor{clCorrect}{HTML}{2E8B57}
\definecolor{clIncorrect}{HTML}{CD5C5C}
\definecolor{clNoGT}{HTML}{E8963E}
\definecolor{clQuestionable}{HTML}{9B7DC4}
\definecolor{clNearMiss}{HTML}{D4A843}
\definecolor{clWrongSel}{HTML}{B84040}
\definecolor{clCouldNotDet}{HTML}{7B8FA1}
\definecolor{clTotal}{HTML}{6B7B8D}

\fill[clTotal, opacity=0.75, rounded corners=1pt] (0.000,0.000) rectangle (1.800,2.750);

\fill[clCorrect, opacity=0.75, rounded corners=1pt] (4.500,0.713) rectangle (6.300,3.278);
\fill[clIncorrect, opacity=0.75, rounded corners=1pt] (4.500,-0.528) rectangle (6.300,0.463);

\fill[clNoGT, opacity=0.75, rounded corners=1pt] (9.000,0.230) rectangle (10.800,0.780);
\fill[clWrongSel, opacity=0.75, rounded corners=1pt] (9.000,-0.845) rectangle (10.800,-0.020);

\fill[clCouldNotDet, opacity=0.75, rounded corners=1pt] (13.500,-1.616) rectangle (15.300,-1.341);
\fill[clWrongSel, opacity=0.75, rounded corners=1pt] (13.500,-1.091) rectangle (15.300,-0.614);
\fill[clNearMiss, opacity=0.75, rounded corners=1pt] (13.500,-0.364) rectangle (15.300,0.024);
\fill[clQuestionable, opacity=0.75, rounded corners=1pt] (13.500,0.274) rectangle (15.300,0.751);

\fill[clCorrect, opacity=0.30] (1.800,0.767) .. controls (3.150,0.767) and (3.150,0.713) .. (4.500,0.713) -- (4.500,3.278) .. controls (3.150,3.278) and (3.150,2.750) .. (1.800,2.750) -- cycle;
\fill[clIncorrect, opacity=0.30] (1.800,0.000) .. controls (3.150,0.000) and (3.150,-0.528) .. (4.500,-0.528) -- (4.500,0.463) .. controls (3.150,0.463) and (3.150,0.767) .. (1.800,0.767) -- cycle;

\fill[clNoGT, opacity=0.30] (6.300,0.067) .. controls (7.650,0.067) and (7.650,0.230) .. (9.000,0.230) -- (9.000,0.780) .. controls (7.650,0.780) and (7.650,0.463) .. (6.300,0.463) -- cycle;
\fill[clWrongSel, opacity=0.30] (6.300,-0.528) .. controls (7.650,-0.528) and (7.650,-0.845) .. (9.000,-0.845) -- (9.000,-0.020) .. controls (7.650,-0.020) and (7.650,0.067) .. (6.300,0.067) -- cycle;

\fill[clCouldNotDet, opacity=0.30] (10.800,-0.845) .. controls (12.150,-0.845) and (12.150,-1.616) .. (13.500,-1.616) -- (13.500,-1.341) .. controls (12.150,-1.341) and (12.150,-0.705) .. (10.800,-0.705) -- cycle;
\fill[clWrongSel, opacity=0.30] (10.800,-0.705) .. controls (12.150,-0.705) and (12.150,-1.091) .. (13.500,-1.091) -- (13.500,-0.614) .. controls (12.150,-0.614) and (12.150,-0.462) .. (10.800,-0.462) -- cycle;
\fill[clNearMiss, opacity=0.30] (10.800,-0.462) .. controls (12.150,-0.462) and (12.150,-0.364) .. (13.500,-0.364) -- (13.500,0.024) .. controls (12.150,0.024) and (12.150,-0.263) .. (10.800,-0.263) -- cycle;
\fill[clQuestionable, opacity=0.30] (10.800,-0.263) .. controls (12.150,-0.263) and (12.150,0.274) .. (13.500,0.274) -- (13.500,0.751) .. controls (12.150,0.751) and (12.150,-0.020) .. (10.800,-0.020) -- cycle;

\node[anchor=east, font=\small] at (-0.150,1.375) {\textbf{100} fix commits};
\node[anchor=south, font=\small, text=clCorrect!90!black] at (5.400,3.378) {\textbf{87} correct};
\node[anchor=north, font=\small, text=clIncorrect!90!black] at (5.400,-0.628) {\textbf{13} incorrect};
\node[anchor=south, font=\small, text=clNoGT!90!black, text width=3.5cm, align=center] at (9.900,0.880) {\textbf{4} BIC not in\\[-2pt]candidate set};
\node[anchor=north, font=\small, text=clWrongSel!90!black, text width=3.5cm, align=center] at (9.900,-0.945) {\textbf{9} BIC in candidate\\[-2pt]set, wrong selection};
\node[anchor=west, font=\small, text=clCouldNotDet!90!black, text width=3.5cm] at (15.450,-1.478) {\textbf{1} could not\\[-2pt]determine};
\node[anchor=west, font=\small, text=clWrongSel!90!black, text width=3.5cm] at (15.450,-0.853) {\textbf{3} incorrect\\[-2pt]analysis};
\node[anchor=west, font=\small, text=clNearMiss!90!black, text width=3.5cm] at (15.450,-0.170) {\textbf{2} near miss\\[-2pt]($\leq$3 positions)};
\node[anchor=west, font=\small, text=clQuestionable!90!black, text width=3.5cm] at (15.450,0.513) {\textbf{3} questionable\\[-2pt]ground truth label};

\end{tikzpicture}

%% file: tables/table_cheaper_models.tex
\begin{table*}[t]
\footnotesize
\centering
\caption{RQ4c: Simple-SZZ-Agent with different underlying models and agents. Best results in \textbf{bold}.}
\label{tab:results_cheaper_models}
\begin{tabular}{llcccccc}
\toprule
 & & & & & \multicolumn{3}{c}{\textit{Avg.\ per Fix Commit}} \\
\cmidrule(l){6-8}
\textbf{Agent} & \textbf{Model} & \textbf{Precision} & \textbf{Recall} & \textbf{F1-Score} & \textbf{Cost} & \textbf{Tool Calls} & \textbf{Tokens} \\
\midrule
Claude Code & claude-haiku-4-5 & 0.75 & 0.73 & 0.74 & \$0.09 & 14.2 & 339k \\
Claude Code & claude-sonnet-4-5 & 0.81 & 0.80 & 0.80 & \$0.25 & 16.7 & 290k \\
Claude Code & claude-opus-4-5 & 0.83 & 0.81 & 0.82 & \$0.38 & 14.3 & \textbf{284k} \\
\midrule
OpenHands & claude-haiku-4-5 & 0.80 & 0.79 & 0.79 & \$0.14 & 15.7 & 1.3M \\
OpenHands & claude-sonnet-4-5 & \textbf{0.87} & \textbf{0.86} & \textbf{0.86} & \$0.38 & \textbf{14.0} & 1.0M \\
OpenHands & claude-opus-4-5 & 0.84 & 0.82 & 0.83 & \$0.63 & 16.9 & 1.0M \\
OpenHands & minimax-m2.5 & 0.80 & 0.78 & 0.79 & \textbf{\$0.04} & 20.0 & 1.3M \\
OpenHands & glm-5 & 0.79 & 0.78 & 0.78 & \$0.26 & 17.8 & 754k \\
\bottomrule
\end{tabular}
\end{table*}

%% file: sections/threats_to_validity.tex
\section{Threats to Validity}
\label{sec:threats_to_validity}
We discuss threats to the internal, external, and construct validity of our study.

\subsection{Internal Validity}

\boldpar{Sample Size} Budget constraints limit our evaluation to 75--243 fix commits per dataset, which may not capture the full distribution of bug types and fix patterns. We mitigate this by using fixed random seeds for sampling, reporting statistical significance tests and effect sizes for all comparisons with previous approaches, and providing all sampling scripts for reproducibility.

\boldpar{Prompt Sensitivity} The instructions given to the agent may influence its behavior and effectiveness. We did not perform systematic prompt optimization; our prompts are straightforward task descriptions and are released with our artifact.

\subsection{External Validity}

\boldpar{Language and Project Scope} Our evaluation covers C/C++ and Java projects across five datasets derived from the Linux kernel, GitHub repositories, and the Apache ecosystem. Results may not generalize to other languages or project types. We selected these datasets because they are the most popular in SZZ research and allow direct comparison with prior work.

\boldpar{LLM Evolution} Our results depend on current LLM capabilities and future models may yield different results. Our evaluation across six models and two agent frameworks~(\S\ref{sec:simple-szz-agent:experiments:rq4c}) demonstrates that the approach is not tied to a specific model.

\boldpar{Generalizability to Novel Projects} None of our experiments ensure that the bug-introducing commits themselves are absent from the training data; even in DS\_LINUX-26, these commits may predate the training cutoff. This is not data leakage, since the information that identifies them as bug-introducing, the \texttt{Fixes:} tag and bug report, is created only after the cutoff and is therefore absent from the training data~(\S\ref{sec:szz-agent:experiments:rq2a}). It does, however, leave open how well SZZ-Agent and Simple-SZZ-Agent transfer to novel projects whose commit history is entirely absent from the training data. Fine-tuning on the target project's commit history could mitigate this, allowing the model to learn project-specific coding patterns and commit conventions before it is applied to identify bug-introducing commits.

\subsection{Construct Validity}

\boldpar{Ground Truth Quality} We rely on developer-annotated bug-introducing commits as ground truth. For DS\_LINUX and DS\_LINUX-26, these annotations come from \texttt{Fixes:} tags added by Linux kernel developers, which may occasionally be incorrect or incomplete. For DS\_GITHUB-c and DS\_GITHUB-j, the annotations were constructed by identifying fix commits where developers explicitly reference bug-introducing commits, followed by manual filtering~\citep{rosa_2021_ds_github}. For DS\_APACHE, the bug-introducing commits were manually collected and verified by \citet{wen_2019_ds_apache}. Inaccuracies in these labels would affect all approaches equally, but could bias absolute performance estimates.

\boldpar{Benchmark Validity} Risse et al.~\cite{risse_2024_uncovering, risse_2025_topscorewrongexam} showed that benchmarks in machine learning for vulnerability detection admit top scores without the capability they claim to measure: the samples lack the context required to decide the label, and classifiers achieve high scores through spurious correlations. Our evaluation seems to be subject to the same critique: To identify the bug-introducing commit, the annotating developers may have drawn on information beyond the provided context, such as bug reports or runtime behavior, leaving some labels undecidable from the input alone, and the agents may exploit superficial cues such as matching removed lines~(\S\ref{sec:simple-szz-agent:experiments:rq4a}) rather than understanding the bug. However, neither threat invalidates our claims. We do not claim that the agents understand bugs, but that they identify the developer-annotated bug-introducing commit more accurately than all previous approaches under identical inputs and labels. Unlike function-level vulnerability detection, our benchmark task is not a proxy for a different capability but the deployment task itself: identifying the developer-annotated commit is itself the task in practice.

\boldpar{Single Bug-Introducing Commit Assumption} Our evaluation metrics assume that each fix commit has a well-defined set of bug-introducing commits. In practice, bugs can result from complex interactions across multiple commits, and the notion of a single introducing commit may oversimplify the true causal history. This limitation applies to all SZZ approaches and is inherited from the standard evaluation methodology~\citep{Lyu_2024_Linux_Kernel_Dataset}.

%% file: tables/table_effectiveness_file_combined.tex
\begin{table*}[t!]
\scriptsize
\setlength{\tabcolsep}{3pt}
\renewcommand{\arraystretch}{0.9}
\centering
\caption{F1-score split by file-level overlap between fix commit and bug-introducing commits. The agents only search file-overlap candidates (N/A without overlap). Best F1 per column in \textbf{bold}; entry counts in headers.}
\label{tab:results_file_combined}
\begin{tabular}{lcccccccc}
\toprule
 & \multicolumn{2}{c}{\textbf{DS\_LINUX}} & \multicolumn{2}{c}{\textbf{DS\_GITHUB-c}} & \multicolumn{2}{c}{\textbf{DS\_GITHUB-j}} & \multicolumn{2}{c}{\textbf{DS\_APACHE}} \\
\cmidrule(lr){2-3} \cmidrule(lr){4-5} \cmidrule(lr){6-7} \cmidrule(lr){8-9}
\textbf{Approach} & \textbf{With} ($n=189$) & \textbf{Without} ($n=11$) & \textbf{With} ($n=90$) & \textbf{Without} ($n=10$) & \textbf{With} ($n=66$) & \textbf{Without} ($n=7$) & \textbf{With} ($n=211$) & \textbf{Without} ($n=32$) \\
\midrule
SZZ & 0.57 & 0.00 & 0.57 & \textbf{0.10} & 0.63 & 0.00 & 0.55 & 0.01 \\
AG-SZZ & 0.55 & 0.00 & 0.58 & \textbf{0.10} & 0.56 & 0.00 & 0.41 & 0.02 \\
L-SZZ & 0.46 & 0.00 & 0.42 & 0.00 & 0.46 & 0.00 & 0.32 & 0.00 \\
R-SZZ & 0.50 & 0.00 & 0.64 & \textbf{0.10} & 0.51 & 0.00 & 0.43 & \textbf{0.03} \\
MA-SZZ & 0.53 & 0.00 & 0.58 & 0.07 & 0.54 & 0.00 & 0.40 & 0.02 \\
RA-SZZ & 0.53 & 0.00 & 0.58 & 0.07 & 0.46 & 0.00 & 0.33 & 0.02 \\
V-SZZ & 0.52 & 0.00 & 0.57 & \textbf{0.10} & 0.61 & 0.00 & 0.54 & 0.01 \\
LLM4SZZ & 0.68 & 0.00 & 0.69 & \textbf{0.10} & 0.60 & 0.00 & 0.59 & 0.00 \\
\midrule
\rowcolor{gray!20} \textbf{SZZ-Agent} & 0.81 & N/A & 0.82 & N/A & 0.74 & N/A & 0.68 & N/A \\
\rowcolor{gray!20} \textbf{Simple-SZZ-Agent} & \textbf{0.86} & N/A & \textbf{0.83} & N/A & \textbf{0.82} & N/A & \textbf{0.75} & N/A \\
\bottomrule
\end{tabular}
\end{table*}

%% file: tables/table_effectiveness_function_combined.tex
\begin{table*}[t!]
\scriptsize
\setlength{\tabcolsep}{3pt}
\renewcommand{\arraystretch}{0.9}
\centering
\caption{F1-score split by function-level overlap between fix commit and bug-introducing commits. Best F1 per column in \textbf{bold}; entry counts in headers.}
\label{tab:results_function_combined}
\begin{tabular}{lcccccccc}
\toprule
 & \multicolumn{2}{c}{\textbf{DS\_LINUX}} & \multicolumn{2}{c}{\textbf{DS\_GITHUB-c}} & \multicolumn{2}{c}{\textbf{DS\_GITHUB-j}} & \multicolumn{2}{c}{\textbf{DS\_APACHE}} \\
\cmidrule(lr){2-3} \cmidrule(lr){4-5} \cmidrule(lr){6-7} \cmidrule(lr){8-9}
\textbf{Approach} & \textbf{With} ($n=93$) & \textbf{Without} ($n=107$) & \textbf{With} ($n=70$) & \textbf{Without} ($n=30$) & \textbf{With} ($n=49$) & \textbf{Without} ($n=24$) & \textbf{With} ($n=149$) & \textbf{Without} ($n=94$) \\
\midrule
SZZ & 0.62 & 0.48 & 0.58 & 0.40 & 0.62 & 0.44 & 0.56 & 0.35 \\
AG-SZZ & 0.62 & 0.44 & 0.59 & 0.40 & 0.55 & 0.42 & 0.43 & 0.23 \\
L-SZZ & 0.48 & 0.40 & 0.39 & 0.37 & 0.42 & 0.40 & 0.32 & 0.20 \\
R-SZZ & 0.57 & 0.39 & 0.65 & 0.43 & 0.48 & 0.44 & 0.48 & 0.22 \\
MA-SZZ & 0.58 & 0.43 & 0.59 & 0.39 & 0.52 & 0.42 & 0.44 & 0.22 \\
RA-SZZ & 0.58 & 0.43 & 0.59 & 0.39 & 0.46 & 0.34 & 0.35 & 0.20 \\
V-SZZ & 0.62 & 0.38 & 0.58 & 0.40 & 0.62 & 0.42 & 0.54 & 0.34 \\
LLM4SZZ & 0.76 & 0.54 & 0.68 & 0.50 & 0.62 & 0.40 & 0.60 & 0.38 \\
\midrule
\rowcolor{gray!20} \textbf{SZZ-Agent} & 0.83 & 0.71 & \textbf{0.84} & 0.53 & 0.80 & 0.40 & 0.70 & 0.44 \\
\rowcolor{gray!20} \textbf{Simple-SZZ-Agent} & \textbf{0.89} & \textbf{0.75} & 0.83 & \textbf{0.57} & \textbf{0.86} & \textbf{0.52} & \textbf{0.76} & \textbf{0.47} \\
\bottomrule
\end{tabular}
\end{table*}

%% file: tables/table_cheaper_models_split.tex
\begin{table*}[t!]
\scriptsize
\setlength{\tabcolsep}{3pt}
\renewcommand{\arraystretch}{0.9}
\centering
\caption{Average effort per fix commit of Simple-SZZ-Agent with different models and agents, split by successful vs.\ unsuccessful predictions. Errored entries are excluded. Lowest values per column in \textbf{bold}.}
\label{tab:results_cheaper_models_split}
\begin{tabular}{llcccccccc}
\toprule
 & & \multicolumn{4}{c}{\textit{Successful (Avg.\ per Fix Commit)}} & \multicolumn{4}{c}{\textit{Unsuccessful (Avg.\ per Fix Commit)}} \\
\cmidrule(lr){3-6} \cmidrule(lr){7-10}
\textbf{Agent} & \textbf{Model} & \textbf{n} & \textbf{Cost} & \textbf{Tool Calls} & \textbf{Tokens} & \textbf{n} & \textbf{Cost} & \textbf{Tool Calls} & \textbf{Tokens} \\
\midrule
Claude Code & claude-haiku-4-5 & 75 & \$0.08 & \textbf{13.2} & 295k & 21 & \$0.13 & 17.8 & 494k \\
Claude Code & claude-sonnet-4-5 & 81 & \$0.25 & 16.5 & \textbf{273k} & 15 & \$0.31 & 17.7 & 383k \\
Claude Code & claude-opus-4-5 & 83 & \$0.38 & 14.0 & 281k & 13 & \$0.46 & 16.3 & \textbf{306k} \\
\midrule
OpenHands & claude-haiku-4-5 & 80 & \$0.14 & 14.1 & 1.1M & 16 & \$0.22 & 23.5 & 2.2M \\
OpenHands & claude-sonnet-4-5 & 87 & \$0.39 & 13.7 & 963k & 9 & \$0.52 & 16.7 & 1.6M \\
OpenHands & claude-opus-4-5 & 84 & \$0.62 & 16.3 & 919k & 12 & \$0.88 & 21.0 & 1.6M \\
OpenHands & minimax-m2.5 & 80 & \textbf{\$0.04} & 19.4 & 1.2M & 15 & \textbf{\$0.06} & 23.5 & 1.6M \\
OpenHands & glm-5 & 79 & \$0.31 & 19.4 & 737k & 15 & \$0.16 & \textbf{9.4} & 440k \\
\bottomrule
\end{tabular}
\end{table*}

%% file: sections/conclusion.tex
\section{Conclusion}
\label{sec:conclusion}
We introduced SZZ-Agent, an agentic approach that uses binary search over source code versions to identify bug-introducing commits, outperforming all previous approaches by at least 9 percentage points in F1-score on four established datasets. Through five ablation studies, we established that this improvement stems from the agentic approach itself, not from data leakage or a more capable LLM backbone.

During our ablations, we discovered that binary search is unnecessary: Simple-SZZ-Agent, which gives a single agent the full candidate set, matches or outperforms SZZ-Agent on all datasets. It distills the fix commit into short greppable patterns to search the candidate set, explaining why cost and performance remain largely independent of the candidate set size. Simple-SZZ-Agent generalizes across agent frameworks and LLM backbones, and even cheap configurations outperform all previous non-agentic approaches.

Beyond identifying bug-introducing commits, the agent's ability to compress bug knowledge into short patterns opens avenues for future work in root cause understanding and program repair. We believe that agents, equipped with simple developer tools like grep, offer a powerful and general strategy for navigating large search spaces in software engineering---and that the problems we studied here are only the beginning.\newpage

%% file: sections/appendix.tex
\section{Appendix}

\subsection{Results Split by Fix--BIC Overlap}
\label{app:overlap}

\boldpar{File-Level Overlap} Table~\ref{tab:results_file_combined} splits each dataset by whether at least one true bug-introducing commit modifies a file that the fix commit also modifies. In 5.5\% (DS\_LINUX) to 13.2\% (DS\_APACHE) of cases, there is no such overlap. Since SZZ-Agent and Simple-SZZ-Agent collect their candidate commits from the histories of the files modified by the fix, they cannot identify the true bug-introducing commit in these cases by construction; in Tables~\ref{tab:results} and~\ref{tab:agent_comparison}, they count as failures. The baselines, whose candidate sets are not restricted in this way, also fail almost completely on this subset (F1 $\leq$ 0.10).

\boldpar{Function-Level Overlap} Table~\ref{tab:results_function_combined} splits each dataset by whether the fix commit and at least one true bug-introducing commit modify the same function. All approaches perform substantially better when the modified functions overlap; for example, Simple-SZZ-Agent achieves 0.89 vs.\ 0.75 F1-score on DS\_LINUX. Simple-SZZ-Agent outperforms all baselines on both subsets of every dataset, but cases without function-level overlap remain substantially harder for every approach.

\subsection{Effort Split by Prediction Outcome}
\label{app:effort-split}

Table~\ref{tab:results_cheaper_models_split} splits the average effort per fix commit of each RQ4c configuration (Table~\ref{tab:results_cheaper_models}) into successful and unsuccessful predictions. In all configurations except OpenHands with glm-5, unsuccessful predictions consume more cost, tool calls, and tokens than successful ones, suggesting that the agents search longer before settling on an incorrect prediction.

%% file: sections/data_availability.tex
\section*{Data Availability Statement}
To ensure the reproducibility of our results and to provide transparency in our research, we have made all related scripts and data publicly available in our artifact: \url{\artifacturl}. Detailed documentation regarding the artifact structure, environment setup, and instructions on how to reproduce the experiments is provided in the \texttt{README.md} file located in the root directory of the artifact.

%% file: sections/acknowledgements.tex
\begin{acks}
This research is funded by the European Union. Views and opinions expressed are
however those of the author(s) only and do not necessarily reflect those of the
European Union or the European Research Council Executive Agency. Neither the
European Union nor the granting authority can be held responsible for them.
This work is supported by ERC grant (Project AT\_SCALE, 101179366).
\end{acks}

%% file: references.bib
@article{Lyu_2024_Linux_Kernel_Dataset,
author = {Lyu, Yunbo and Kang, Hong Jin and Widyasari, Ratnadira and Lawall, Julia and Lo, David},
title = {Evaluating SZZ Implementations: An Empirical Study on the Linux Kernel},
year = {2024},
issue_date = {Sept. 2024},
publisher = {IEEE Press},
volume = {50},
number = {9},
issn = {0098-5589},
url = {https://doi.org/10.1109/TSE.2024.3406718},
doi = {10.1109/TSE.2024.3406718},
journal = {IEEE Trans. Softw. Eng.},
month = sep,
pages = {2219–2239},
numpages = {21}
}

@article{sliwerski_2005_original_szz,
author = {Śliwerski, Jacek and Zimmermann, Thomas and Zeller, Andreas},
title = {When do changes induce fixes?},
year = {2005},
issue_date = {July 2005},
publisher = {Association for Computing Machinery},
address = {New York, NY, USA},
volume = {30},
number = {4},
issn = {0163-5948},
url = {https://doi.org/10.1145/1082983.1083147},
doi = {10.1145/1082983.1083147},
journal = {SIGSOFT Softw. Eng. Notes},
month = may,
pages = {1–5},
numpages = {5}
}

@inproceedings{sunghun_2006_agszz,
author = {Kim, Sunghun and Zimmermann, Thomas and Pan, Kai and Whitehead, E. James Jr.},
title = {Automatic Identification of Bug-Introducing Changes},
year = {2006},
isbn = {0769525792},
publisher = {IEEE Computer Society},
address = {USA},
url = {https://doi.org/10.1109/ASE.2006.23},
doi = {10.1109/ASE.2006.23},
booktitle = {Proceedings of the 21st IEEE/ACM International Conference on Automated Software Engineering},
pages = {81–90},
numpages = {10},
series = {ASE '06}
}

@online{anthropic_models_overview,
  author       = {{Anthropic}},
  title        = {Models Overview},
  year         = {2024},
  url          = {https://platform.claude.com/docs/en/about-claude/models/overview},
  urldate      = {2024-02-02},
  organization = {Anthropic PBC},
  note         = {Claude API Documentation}
}

@article{davies_2014_lszz_rszz,
author = {Davies, Steven and Roper, Marc and Wood, Murray},
title = {Comparing text-based and dependence-based approaches for determining the origins of bugs},
journal = {Journal of Software: Evolution and Process},
volume = {26},
number = {1},
pages = {107-139},
doi = {https://doi.org/10.1002/smr.1619},
url = {https://onlinelibrary.wiley.com/doi/abs/10.1002/smr.1619},
eprint = {https://onlinelibrary.wiley.com/doi/pdf/10.1002/smr.1619},
year = {2014}
}

@article{dacosta_2017_maszz,
author = {da Costa, Daniel Alencar and McIntosh, Shane and Shang, Weiyi and Kulesza, Uir\'{a} and Coelho, Roberta and Hassan, Ahmed E.},
title = {A Framework for Evaluating the Results of the SZZ Approach for Identifying Bug-Introducing Changes},
year = {2017},
issue_date = {July 2017},
publisher = {IEEE Press},
volume = {43},
number = {7},
issn = {0098-5589},
url = {https://doi.org/10.1109/TSE.2016.2616306},
doi = {10.1109/TSE.2016.2616306},
journal = {IEEE Trans. Softw. Eng.},
month = jul,
pages = {641–657},
numpages = {17}
}

@inproceedings{neto_2018_raszz,
  author={Neto, Edmilson Campos and da Costa, Daniel Alencar and Kulesza, Uirá},
  booktitle={2018 IEEE 25th International Conference on Software Analysis, Evolution and Reengineering (SANER)}, 
  title={The impact of refactoring changes on the SZZ algorithm: An empirical study}, 
  year={2018},
  volume={},
  number={},
  pages={380-390},
  doi={10.1109/SANER.2018.8330225}
}

@inproceedings{tang_2023_neuralszz,
author = {Tang, Lingxiao and Bao, Lingfeng and Xia, Xin and Huang, Zhongdong},
title = {Neural SZZ Algorithm},
year = {2024},
isbn = {9798350329964},
publisher = {IEEE Press},
url = {https://doi.org/10.1109/ASE56229.2023.00037},
doi = {10.1109/ASE56229.2023.00037},
booktitle = {Proceedings of the 38th IEEE/ACM International Conference on Automated Software Engineering},
pages = {1024–1035},
numpages = {12},
location = {Echternach, Luxembourg},
series = {ASE '23}
}

@article{tang_2025_llm4szz,
author = {Tang, Lingxiao and Liu, Jiakun and Liu, Zhongxin and Yang, Xiaohu and Bao, Lingfeng},
title = {LLM4SZZ: Enhancing SZZ Algorithm with Context-Enhanced Assessment on Large Language Models},
year = {2025},
issue_date = {July 2025},
publisher = {Association for Computing Machinery},
address = {New York, NY, USA},
volume = {2},
number = {ISSTA},
url = {https://doi.org/10.1145/3728885},
doi = {10.1145/3728885},
journal = {Proc. ACM Softw. Eng.},
month = jun,
articleno = {ISSTA016},
numpages = {23}
}

@inproceedings{bao_2022_vszz,
author = {Bao, Lingfeng and Xia, Xin and Hassan, Ahmed E. and Yang, Xiaohu},
title = {V-SZZ: automatic identification of version ranges affected by CVE vulnerabilities},
year = {2022},
isbn = {9781450392211},
publisher = {Association for Computing Machinery},
address = {New York, NY, USA},
url = {https://doi.org/10.1145/3510003.3510113},
doi = {10.1145/3510003.3510113},
booktitle = {Proceedings of the 44th International Conference on Software Engineering},
pages = {2352–2364},
numpages = {13},
location = {Pittsburgh, Pennsylvania},
series = {ICSE '22}
}

@inproceedings{rosa_2021_ds_github,
author = {Rosa, Giovanni and Pascarella, Luca and Scalabrino, Simone and Tufano, Rosalia and Bavota, Gabriele and Lanza, Michele and Oliveto, Rocco},
title = {Evaluating SZZ Implementations Through a Developer-informed Oracle},
year = {2021},
isbn = {9781450390859},
publisher = {IEEE Press},
url = {https://doi.org/10.1109/ICSE43902.2021.00049},
doi = {10.1109/ICSE43902.2021.00049},
booktitle = {Proceedings of the 43rd International Conference on Software Engineering},
pages = {436–447},
numpages = {12},
location = {Madrid, Spain},
series = {ICSE '21}
}

@article{rosa_2023_szzvariants,
  title     = {A comprehensive evaluation of SZZ Variants through a developer-informed oracle},
  author    = {Rosa, Giovanni and Pascarella, Luca and Scalabrino, Simone and Tufano, Rosalia and Bavota, Gabriele and Lanza, Michele and Oliveto, Rocco},
  journal   = {Journal of Systems and Software},
  volume    = {202},
  pages     = {111729},
  year      = {2023},
  publisher = {Elsevier},
  doi       = {10.1016/j.jss.2023.111729}
}

@online{anthropic_claude_code,
  author       = {{Anthropic}},
  title        = {Claude Code},
  year         = {2025},
  url          = {https://github.com/anthropics/claude-code},
  urldate      = {2026-03-09}
}

@online{anthropic_claude_code_tools,
  author       = {{Anthropic}},
  title        = {Claude Code Tools Reference},
  year         = {2025},
  url          = {https://code.claude.com/docs/en/tools-reference},
  urldate      = {2026-03-20}
}

@article{wilcoxon_1945,
  author  = {Wilcoxon, Frank},
  title   = {Individual Comparisons by Ranking Methods},
  journal = {Biometrics Bulletin},
  volume  = {1},
  number  = {6},
  pages   = {80--83},
  year    = {1945},
  doi     = {10.2307/3001968}
}

@article{kerby_2014,
  author  = {Kerby, Dave S.},
  title   = {The Simple Difference Formula: An Approach to Teaching Nonparametric Correlation},
  journal = {Comprehensive Psychology},
  volume  = {3},
  pages   = {11.IT.3.1},
  year    = {2014},
  doi     = {10.2466/11.IT.3.1}
}

@article{risse_2025_topscorewrongexam,
author = {Risse, Niklas and Liu, Jing and B\"{o}hme, Marcel},
title = {Top Score on the Wrong Exam: On Benchmarking in Machine Learning for Vulnerability Detection},
year = {2025},
issue_date = {July 2025},
publisher = {Association for Computing Machinery},
address = {New York, NY, USA},
volume = {2},
number = {ISSTA},
url = {https://doi.org/10.1145/3728887},
doi = {10.1145/3728887},
journal = {Proc. ACM Softw. Eng.},
month = jun,
articleno = {ISSTA018},
numpages = {23}
}

@misc{jin_2025_agents_in_se_survey,
      title={From LLMs to LLM-based Agents for Software Engineering: A Survey of Current, Challenges and Future}, 
      author={Haolin Jin and Linghan Huang and Haipeng Cai and Jun Yan and Bo Li and Huaming Chen},
      year={2025},
      eprint={2408.02479},
      archivePrefix={arXiv},
      primaryClass={cs.SE},
      url={https://arxiv.org/abs/2408.02479},
      doi={10.48550/arXiv.2408.02479}, 
}

@article{wang_2024_openhands,
  author  = {Wang, Xingyao and Li, Boxuan and Song, Yufan and Xu, Frank F. and Tang, Xiangru and Zhuge, Mingchen and Pan, Jiayi and Song, Yueqi and Li, Bowen and Singh, Jaskirat and Tran, Hoang H. and Li, Fuqiang and Ma, Ren and Zheng, Mingzhang and Qian, Bill and Shao, Yanjun and Muennighoff, Niklas and Zhang, Yizhe and Hui, Binyuan and Lin, Junyang and Brennan, Robert and Peng, Hao and Ji, Heng and Neubig, Graham},
  title   = {OpenHands: An Open Platform for AI Software Developers as Generalist Agents},
  journal = {arXiv},
  year    = {2024},
  doi     = {10.48550/arxiv.2407.16741}
}

@inproceedings{schick_2023_toolformer,
author = {Schick, Timo and Dwivedi-Yu, Jane and Dess\'{\i}, Roberto and Raileanu, Roberta and Lomeli, Maria and Hambro, Eric and Zettlemoyer, Luke and Cancedda, Nicola and Scialom, Thomas},
title = {Toolformer: language models can teach themselves to use tools},
year = {2023},
publisher = {Curran Associates Inc.},
address = {Red Hook, NY, USA},
booktitle = {Proceedings of the 37th International Conference on Neural Information Processing Systems},
articleno = {2997},
numpages = {13},
location = {New Orleans, LA, USA},
series = {NIPS '23},
doi = {10.48550/arXiv.2302.04761}
}

@inproceedings{yao_2023_react,
  title = {{ReAct}: Synergizing Reasoning and Acting in Language Models},
  author = {Yao, Shunyu and Zhao, Jeffrey and Yu, Dian and Du, Nan and Shafran, Izhak and Narasimhan, Karthik and Cao, Yuan},
  booktitle = {International Conference on Learning Representations (ICLR) },
  year = {2023},
  html = {https://arxiv.org/abs/2210.03629},
  doi = {10.48550/arXiv.2210.03629},
}

@inproceedings{yang_2024_sweagent,
author = {Yang, John and Jimenez, Carlos E. and Wettig, Alexander and Lieret, Kilian and Yao, Shunyu and Narasimhan, Karthik and Press, Ofir},
title = {SWE-agent: agent-computer interfaces enable automated software engineering},
year = {2024},
isbn = {9798331314385},
publisher = {Curran Associates Inc.},
address = {Red Hook, NY, USA},
booktitle = {Proceedings of the 38th International Conference on Neural Information Processing Systems},
articleno = {1601},
numpages = {125},
location = {Vancouver, BC, Canada},
series = {NIPS '24},
doi = {10.48550/arXiv.2405.15793}
}

@inproceedings{zhang_2024_autocoderover,
author = {Zhang, Yuntong and Ruan, Haifeng and Fan, Zhiyu and Roychoudhury, Abhik},
title = {AutoCodeRover: Autonomous Program Improvement},
year = {2024},
isbn = {9798400706127},
publisher = {Association for Computing Machinery},
address = {New York, NY, USA},
url = {https://doi.org/10.1145/3650212.3680384},
doi = {10.1145/3650212.3680384},
booktitle = {Proceedings of the 33rd ACM SIGSOFT International Symposium on Software Testing and Analysis},
pages = {1592–1604},
numpages = {13},
location = {Vienna, Austria},
series = {ISSTA 2024}
}

@online{linux_commit_1d66c3f2b8c0,
  author = {{Linux Kernel}},
  title  = {drm/amd/display: use udelay rather than fsleep},
  year   = {2025},
  url    = {https://github.com/torvalds/linux/commit/1d66c3f2b8c0}
}

@online{linux_commit_01f60348d8fb,
  author = {{Linux Kernel}},
  title  = {drm/amd/display: Fix `failed to blank crtc!'},
  year   = {2025},
  url    = {https://github.com/torvalds/linux/commit/01f60348d8fb}
}

@online{linux_commit_ba1e9421cf1a,
  author = {{Linux Kernel}},
  title  = {net/smc: fix one NULL pointer dereference in smc\_ib\_is\_sg\_need\_sync()},
  year   = {2025},
  url    = {https://github.com/torvalds/linux/commit/ba1e9421cf1a}
}

@online{linux_commit_0ef69e788411,
  author = {{Linux Kernel}},
  title  = {net/smc: optimize for smc\_sndbuf\_sync\_sg\_for\_device and smc\_rmb\_sync\_sg\_for\_cpu},
  year   = {2022},
  url    = {https://github.com/torvalds/linux/commit/0ef69e788411}
}

@online{linux_commit_b4789aac9d34,
  author = {{Linux Kernel}},
  title  = {drm/msm/a6xx: Fix GMU firmware parser},
  year   = {2025},
  url    = {https://github.com/torvalds/linux/commit/b4789aac9d34}
}

@online{linux_commit_c6ed04f856a4,
  author = {{Linux Kernel}},
  title  = {drm/msm/a6xx: A640/A650 GMU firmware path},
  year   = {2020},
  url    = {https://github.com/torvalds/linux/commit/c6ed04f856a4}
}

@InProceedings{gu_2024_cruxeval,
  title = 	 {{CRUXE}val: A Benchmark for Code Reasoning, Understanding and Execution},
  author =       {Gu, Alex and Roziere, Baptiste and Leather, Hugh James and Solar-Lezama, Armando and Synnaeve, Gabriel and Wang, Sida},
  booktitle = 	 {Proceedings of the 41st International Conference on Machine Learning},
  pages = 	 {16568--16621},
  year = 	 {2024},
  editor = 	 {Salakhutdinov, Ruslan and Kolter, Zico and Heller, Katherine and Weller, Adrian and Oliver, Nuria and Scarlett, Jonathan and Berkenkamp, Felix},
  volume = 	 {235},
  series = 	 {Proceedings of Machine Learning Research},
  month = 	 {21--27 Jul},
  publisher =    {PMLR},
  url = 	 {https://proceedings.mlr.press/v235/gu24c.html},
  doi = 	 {10.48550/arXiv.2401.03065}
}

@INPROCEEDINGS{chen_2025_how_far_are_we,
  author={Chen, Xingchu and Liu, Chengwei and Cao, Jialun and Xiao, Yang and Cai, Xinyue and Li, Yeting and Shi, Jingyi and Sun, Tianqi and Chen, Haiming and Huo, Wei},
  booktitle={2025 40th IEEE/ACM International Conference on Automated Software Engineering (ASE)}, 
  title={Vulnerability-Affected Versions Identification: How Far Are We?}, 
  year={2025},
  volume={},
  number={},
  pages={2970-2982},
  doi={10.1109/ASE63991.2025.00244}}

@inproceedings{wen_2019_ds_apache,
author = {Wen, Ming and Wu, Rongxin and Liu, Yepang and Tian, Yongqiang and Xie, Xuan and Cheung, Shing-Chi and Su, Zhendong},
title = {Exploring and exploiting the correlations between bug-inducing and bug-fixing commits},
year = {2019},
isbn = {9781450355728},
publisher = {Association for Computing Machinery},
address = {New York, NY, USA},
url = {https://doi.org/10.1145/3338906.3338962},
doi = {10.1145/3338906.3338962},
booktitle = {Proceedings of the 2019 27th ACM Joint Meeting on European Software Engineering Conference and Symposium on the Foundations of Software Engineering},
pages = {326--337},
numpages = {12},
location = {Tallinn, Estonia},
series = {ESEC/FSE 2019}
}

@inproceedings{risse_2024_uncovering,
author = {Risse, Niklas and B\"{o}hme, Marcel},
title = {Uncovering the limits of machine learning for automatic vulnerability detection},
year = {2024},
isbn = {978-1-939133-44-1},
publisher = {USENIX Association},
address = {USA},
booktitle = {Proceedings of the 33rd USENIX Conference on Security Symposium},
articleno = {238},
numpages = {18},
location = {Philadelphia, PA, USA},
series = {SEC '24}
}
